\pdfoutput=1 % forces latex to use pdflatex
\documentclass{llncs}

\usepackage[dvips]{rotating}
\usepackage{RSalgorithmic}
\usepackage{algorithm}
\usepackage{amscd}
\usepackage{amsfonts}
\usepackage{amsmath}
\usepackage{booktabs}
\usepackage{color}
\usepackage{epsfig}
\usepackage{graphicx}
\usepackage{latexsym}
\usepackage{marginnote}
\usepackage{multirow}
\usepackage{paralist}
\usepackage{placeins}
\usepackage{rotate} 
\usepackage{times} 
\usepackage{url}
\usepackage{verbatim} 
\usepackage{wrapfig}
\usepackage{xspace}
\usepackage{datetime}
\usepackage[FIGBOTCAP,TABBOTCAP,center,nooneline]{subfigure}
\usepackage{xcolor,colortbl}
\usepackage{varwidth}

\definecolor {snow}                {rgb}{1.00,0.98,0.98}
\definecolor {ghostwhite}          {rgb}{0.97,0.97,1.00}
\definecolor {whitesmoke}          {rgb}{0.96,0.96,0.96}
\definecolor {gainsboro}           {rgb}{0.86,0.86,0.86}
\definecolor {floralwhite}         {rgb}{1.00,0.98,0.94}
\definecolor {oldlace}             {rgb}{0.99,0.96,0.90}
\definecolor {linen}               {rgb}{0.98,0.94,0.90}
\definecolor {antiquewhite}        {rgb}{0.98,0.92,0.84}
\definecolor {papayawhip}          {rgb}{1.00,0.94,0.84}
\definecolor {blanchedalmond}      {rgb}{1.00,0.92,0.80}
\definecolor {bisque}              {rgb}{1.00,0.89,0.77}
\definecolor {peachpuff}           {rgb}{1.00,0.85,0.73}
\definecolor {navajowhite}         {rgb}{1.00,0.87,0.68}
\definecolor {moccasin}            {rgb}{1.00,0.89,0.71}
\definecolor {cornsilk}            {rgb}{1.00,0.97,0.86}
\definecolor {ivory}               {rgb}{1.00,1.00,0.94}
\definecolor {lemonchiffon}        {rgb}{1.00,0.98,0.80}
\definecolor {seashell}            {rgb}{1.00,0.96,0.93}
\definecolor {honeydew}            {rgb}{0.94,1.00,0.94}
\definecolor {mintcream}           {rgb}{0.96,1.00,0.98}
\definecolor {azure}               {rgb}{0.94,1.00,1.00}
\definecolor {aliceblue}           {rgb}{0.94,0.97,1.00}
\definecolor {lavender}            {rgb}{0.90,0.90,0.98}
\definecolor {lavenderblush}       {rgb}{1.00,0.94,0.96}
\definecolor {mistyrose}           {rgb}{1.00,0.89,0.88}
\definecolor {white}               {rgb}{1.00,1.00,1.00}
\definecolor {black}               {rgb}{0.00,0.00,0.00}
\definecolor {darkslategray}       {rgb}{0.18,0.31,0.31}
\definecolor {dimgray}             {rgb}{0.41,0.41,0.41}
\definecolor {slategray}           {rgb}{0.44,0.50,0.56}
\definecolor {lightslategray}      {rgb}{0.47,0.53,0.60}
\definecolor {gray}                {rgb}{0.75,0.75,0.75}
\definecolor {lightgrey}           {rgb}{0.83,0.83,0.83}
\definecolor {midnightblue}        {rgb}{0.10,0.10,0.44}
\definecolor {navy}                {rgb}{0.00,0.00,0.50}
\definecolor {cornflowerblue}      {rgb}{0.39,0.58,0.93}
\definecolor {darkslateblue}       {rgb}{0.28,0.24,0.55}
\definecolor {slateblue}           {rgb}{0.42,0.35,0.80}
\definecolor {mediumslateblue}     {rgb}{0.48,0.41,0.93}
\definecolor {lightslateblue}      {rgb}{0.52,0.44,1.00}
\definecolor {mediumblue}          {rgb}{0.00,0.00,0.80}
\definecolor {royalblue}           {rgb}{0.25,0.41,0.88}
\definecolor {blue}                {rgb}{0.00,0.00,1.00}
\definecolor {dodgerblue}          {rgb}{0.12,0.56,1.00}
\definecolor {deepskyblue}         {rgb}{0.00,0.75,1.00}
\definecolor {skyblue}             {rgb}{0.53,0.81,0.92}
\definecolor {lightskyblue}        {rgb}{0.53,0.81,0.98}
\definecolor {steelblue}           {rgb}{0.27,0.51,0.71}
\definecolor {lightsteelblue}      {rgb}{0.69,0.77,0.87}
\definecolor {lightblue}           {rgb}{0.68,0.85,0.90}
\definecolor {powderblue}          {rgb}{0.69,0.88,0.90}
\definecolor {paleturquoise}       {rgb}{0.69,0.93,0.93}
\definecolor {darkturquoise}       {rgb}{0.00,0.81,0.82}
\definecolor {mediumturquoise}     {rgb}{0.28,0.82,0.80}
\definecolor {turquoise}           {rgb}{0.25,0.88,0.82}
\definecolor {cyan}                {rgb}{0.00,1.00,1.00}
\definecolor {lightcyan}           {rgb}{0.88,1.00,1.00}
\definecolor {cadetblue}           {rgb}{0.37,0.62,0.63}
\definecolor {mediumaquamarine}    {rgb}{0.40,0.80,0.67}
\definecolor {aquamarine}          {rgb}{0.50,1.00,0.83}
\definecolor {darkgreen}           {rgb}{0.00,0.39,0.00}
\definecolor {darkolivegreen}      {rgb}{0.33,0.42,0.18}
\definecolor {darkseagreen}        {rgb}{0.56,0.74,0.56}
\definecolor {seagreen}            {rgb}{0.18,0.55,0.34}
\definecolor {mediumseagreen}      {rgb}{0.24,0.70,0.44}
\definecolor {lightseagreen}       {rgb}{0.13,0.70,0.67}
\definecolor {palegreen}           {rgb}{0.60,0.98,0.60}
\definecolor {springgreen}         {rgb}{0.00,1.00,0.50}
\definecolor {lawngreen}           {rgb}{0.49,0.99,0.00}
\definecolor {green}               {rgb}{0.00,1.00,0.00}
\definecolor {chartreuse}          {rgb}{0.50,1.00,0.00}
\definecolor {mediumspringgreen}   {rgb}{0.00,0.98,0.60}
\definecolor {greenyellow}         {rgb}{0.68,1.00,0.18}
\definecolor {limegreen}           {rgb}{0.20,0.80,0.20}
\definecolor {yellowgreen}         {rgb}{0.60,0.80,0.20}
\definecolor {forestgreen}         {rgb}{0.13,0.55,0.13}
\definecolor {olivedrab}           {rgb}{0.42,0.56,0.14}
\definecolor {darkkhaki}           {rgb}{0.74,0.72,0.42}
\definecolor {khaki}               {rgb}{0.94,0.90,0.55}
\definecolor {palegoldenrod}       {rgb}{0.93,0.91,0.67}
\definecolor {lightgoldenrodyellow} {rgb}{0.98,0.98,0.82}
\definecolor {lightyellow}         {rgb}{1.00,1.00,0.88}
\definecolor {yellow}              {rgb}{1.00,1.00,0.00}
\definecolor {gold}                {rgb}{1.00,0.84,0.00}
\definecolor {lightgoldenrod}      {rgb}{0.93,0.87,0.51}
\definecolor {goldenrod}           {rgb}{0.85,0.65,0.13}
\definecolor {darkgoldenrod}       {rgb}{0.72,0.53,0.04}
\definecolor {rosybrown}           {rgb}{0.74,0.56,0.56}
\definecolor {indianred}           {rgb}{0.80,0.36,0.36}
\definecolor {saddlebrown}         {rgb}{0.55,0.27,0.07}
\definecolor {sienna}              {rgb}{0.63,0.32,0.18}
\definecolor {peru}                {rgb}{0.80,0.52,0.25}
\definecolor {burlywood}           {rgb}{0.87,0.72,0.53}
\definecolor {beige}               {rgb}{0.96,0.96,0.86}
\definecolor {wheat}               {rgb}{0.96,0.87,0.70}
\definecolor {sandybrown}          {rgb}{0.96,0.64,0.38}
\definecolor {tan}                 {rgb}{0.82,0.71,0.55}
\definecolor {chocolate}           {rgb}{0.82,0.41,0.12}
\definecolor {firebrick}           {rgb}{0.70,0.13,0.13}
\definecolor {brown}               {rgb}{0.65,0.16,0.16}
\definecolor {darksalmon}          {rgb}{0.91,0.59,0.48}
\definecolor {salmon}              {rgb}{0.98,0.50,0.45}
\definecolor {lightsalmon}         {rgb}{1.00,0.63,0.48}
\definecolor {orange}              {rgb}{1.00,0.65,0.00}
\definecolor {darkorange}          {rgb}{1.00,0.55,0.00}
\definecolor {coral}               {rgb}{1.00,0.50,0.31}
\definecolor {lightcoral}          {rgb}{0.94,0.50,0.50}
\definecolor {tomato}              {rgb}{1.00,0.39,0.28}
\definecolor {orangered}           {rgb}{1.00,0.27,0.00}
\definecolor {red}                 {rgb}{1.00,0.00,0.00}
\definecolor {hotpink}             {rgb}{1.00,0.41,0.71}
\definecolor {deeppink}            {rgb}{1.00,0.08,0.58}
\definecolor {pink}                {rgb}{1.00,0.75,0.80}
\definecolor {lightpink}           {rgb}{1.00,0.71,0.76}
\definecolor {palevioletred}       {rgb}{0.86,0.44,0.58}
\definecolor {maroon}              {rgb}{0.69,0.19,0.38}
\definecolor {mediumvioletred}     {rgb}{0.78,0.08,0.52}
\definecolor {violetred}           {rgb}{0.82,0.13,0.56}
\definecolor {magenta}             {rgb}{1.00,0.00,1.00}
\definecolor {violet}              {rgb}{0.93,0.51,0.93}
\definecolor {plum}                {rgb}{0.87,0.63,0.87}
\definecolor {orchid}              {rgb}{0.85,0.44,0.84}
\definecolor {mediumorchid}        {rgb}{0.73,0.33,0.83}
\definecolor {darkorchid}          {rgb}{0.60,0.20,0.80}
\definecolor {darkviolet}          {rgb}{0.58,0.00,0.83}
\definecolor {blueviolet}          {rgb}{0.54,0.17,0.89}
\definecolor {purple}              {rgb}{0.63,0.13,0.94}
\definecolor {mediumpurple}        {rgb}{0.58,0.44,0.86}
\definecolor {thistle}             {rgb}{0.85,0.75,0.85}
\definecolor {snow2}               {rgb}{0.93,0.91,0.91}
\definecolor {snow3}               {rgb}{0.80,0.79,0.79}
\definecolor {snow4}               {rgb}{0.55,0.54,0.54}
\definecolor {seashell2}           {rgb}{0.93,0.90,0.87}
\definecolor {seashell3}           {rgb}{0.80,0.77,0.75}
\definecolor {seashell4}           {rgb}{0.55,0.53,0.51}
\definecolor {antiquewhite1}       {rgb}{1.00,0.94,0.86}
\definecolor {antiquewhite2}       {rgb}{0.93,0.87,0.80}
\definecolor {antiquewhite3}       {rgb}{0.80,0.75,0.69}
\definecolor {antiquewhite4}       {rgb}{0.55,0.51,0.47}
\definecolor {bisque2}             {rgb}{0.93,0.84,0.72}
\definecolor {bisque3}             {rgb}{0.80,0.72,0.62}
\definecolor {bisque4}             {rgb}{0.55,0.49,0.42}
\definecolor {peachpuff2}          {rgb}{0.93,0.80,0.68}
\definecolor {peachpuff3}          {rgb}{0.80,0.69,0.58}
\definecolor {peachpuff4}          {rgb}{0.55,0.47,0.40}
\definecolor {navajowhite2}        {rgb}{0.93,0.81,0.63}
\definecolor {navajowhite3}        {rgb}{0.80,0.70,0.55}
\definecolor {navajowhite4}        {rgb}{0.55,0.47,0.37}
\definecolor {lemonchiffon2}       {rgb}{0.93,0.91,0.75}
\definecolor {lemonchiffon3}       {rgb}{0.80,0.79,0.65}
\definecolor {lemonchiffon4}       {rgb}{0.55,0.54,0.44}
\definecolor {cornsilk2}           {rgb}{0.93,0.91,0.80}
\definecolor {cornsilk3}           {rgb}{0.80,0.78,0.69}
\definecolor {cornsilk4}           {rgb}{0.55,0.53,0.47}
\definecolor {ivory2}              {rgb}{0.93,0.93,0.88}
\definecolor {ivory3}              {rgb}{0.80,0.80,0.76}
\definecolor {ivory4}              {rgb}{0.55,0.55,0.51}
\definecolor {honeydew2}           {rgb}{0.88,0.93,0.88}
\definecolor {honeydew3}           {rgb}{0.76,0.80,0.76}
\definecolor {honeydew4}           {rgb}{0.51,0.55,0.51}
\definecolor {lavenderblush2}      {rgb}{0.93,0.88,0.90}
\definecolor {lavenderblush3}      {rgb}{0.80,0.76,0.77}
\definecolor {lavenderblush4}      {rgb}{0.55,0.51,0.53}
\definecolor {mistyrose2}          {rgb}{0.93,0.84,0.82}
\definecolor {mistyrose3}          {rgb}{0.80,0.72,0.71}
\definecolor {mistyrose4}          {rgb}{0.55,0.49,0.48}
\definecolor {azure2}              {rgb}{0.88,0.93,0.93}
\definecolor {azure3}              {rgb}{0.76,0.80,0.80}
\definecolor {azure4}              {rgb}{0.51,0.55,0.55}
\definecolor {slateblue1}          {rgb}{0.51,0.44,1.00}
\definecolor {slateblue2}          {rgb}{0.48,0.40,0.93}
\definecolor {slateblue3}          {rgb}{0.41,0.35,0.80}
\definecolor {slateblue4}          {rgb}{0.28,0.24,0.55}
\definecolor {royalblue1}          {rgb}{0.28,0.46,1.00}
\definecolor {royalblue2}          {rgb}{0.26,0.43,0.93}
\definecolor {royalblue3}          {rgb}{0.23,0.37,0.80}
\definecolor {royalblue4}          {rgb}{0.15,0.25,0.55}
\definecolor {blue2}               {rgb}{0.00,0.00,0.93}
\definecolor {blue4}               {rgb}{0.00,0.00,0.55}
\definecolor {dodgerblue2}         {rgb}{0.11,0.53,0.93}
\definecolor {dodgerblue3}         {rgb}{0.09,0.45,0.80}
\definecolor {dodgerblue4}         {rgb}{0.06,0.31,0.55}
\definecolor {steelblue1}          {rgb}{0.39,0.72,1.00}
\definecolor {steelblue2}          {rgb}{0.36,0.67,0.93}
\definecolor {steelblue3}          {rgb}{0.31,0.58,0.80}
\definecolor {steelblue4}          {rgb}{0.21,0.39,0.55}
\definecolor {deepskyblue2}        {rgb}{0.00,0.70,0.93}
\definecolor {deepskyblue3}        {rgb}{0.00,0.60,0.80}
\definecolor {deepskyblue4}        {rgb}{0.00,0.41,0.55}
\definecolor {skyblue1}            {rgb}{0.53,0.81,1.00}
\definecolor {skyblue2}            {rgb}{0.49,0.75,0.93}
\definecolor {skyblue3}            {rgb}{0.42,0.65,0.80}
\definecolor {skyblue4}            {rgb}{0.29,0.44,0.55}
\definecolor {lightskyblue1}       {rgb}{0.69,0.89,1.00}
\definecolor {lightskyblue2}       {rgb}{0.64,0.83,0.93}
\definecolor {lightskyblue3}       {rgb}{0.55,0.71,0.80}
\definecolor {lightskyblue4}       {rgb}{0.38,0.48,0.55}
\definecolor {slategray1}          {rgb}{0.78,0.89,1.00}
\definecolor {slategray2}          {rgb}{0.73,0.83,0.93}
\definecolor {slategray3}          {rgb}{0.62,0.71,0.80}
\definecolor {slategray4}          {rgb}{0.42,0.48,0.55}
\definecolor {lightsteelblue1}     {rgb}{0.79,0.88,1.00}
\definecolor {lightsteelblue2}     {rgb}{0.74,0.82,0.93}
\definecolor {lightsteelblue3}     {rgb}{0.64,0.71,0.80}
\definecolor {lightsteelblue4}     {rgb}{0.43,0.48,0.55}
\definecolor {lightblue1}          {rgb}{0.75,0.94,1.00}
\definecolor {lightblue2}          {rgb}{0.70,0.87,0.93}
\definecolor {lightblue3}          {rgb}{0.60,0.75,0.80}
\definecolor {lightblue4}          {rgb}{0.41,0.51,0.55}
\definecolor {lightcyan2}          {rgb}{0.82,0.93,0.93}
\definecolor {lightcyan3}          {rgb}{0.71,0.80,0.80}
\definecolor {lightcyan4}          {rgb}{0.48,0.55,0.55}
\definecolor {paleturquoise1}      {rgb}{0.73,1.00,1.00}
\definecolor {paleturquoise2}      {rgb}{0.68,0.93,0.93}
\definecolor {paleturquoise3}      {rgb}{0.59,0.80,0.80}
\definecolor {paleturquoise4}      {rgb}{0.40,0.55,0.55}
\definecolor {cadetblue1}          {rgb}{0.60,0.96,1.00}
\definecolor {cadetblue2}          {rgb}{0.56,0.90,0.93}
\definecolor {cadetblue3}          {rgb}{0.48,0.77,0.80}
\definecolor {cadetblue4}          {rgb}{0.33,0.53,0.55}
\definecolor {turquoise1}          {rgb}{0.00,0.96,1.00}
\definecolor {turquoise2}          {rgb}{0.00,0.90,0.93}
\definecolor {turquoise3}          {rgb}{0.00,0.77,0.80}
\definecolor {turquoise4}          {rgb}{0.00,0.53,0.55}
\definecolor {cyan2}               {rgb}{0.00,0.93,0.93}
\definecolor {cyan3}               {rgb}{0.00,0.80,0.80}
\definecolor {cyan4}               {rgb}{0.00,0.55,0.55}
\definecolor {darkslategray1}      {rgb}{0.59,1.00,1.00}
\definecolor {darkslategray2}      {rgb}{0.55,0.93,0.93}
\definecolor {darkslategray3}      {rgb}{0.47,0.80,0.80}
\definecolor {darkslategray4}      {rgb}{0.32,0.55,0.55}
\definecolor {aquamarine2}         {rgb}{0.46,0.93,0.78}
\definecolor {aquamarine4}         {rgb}{0.27,0.55,0.45}
\definecolor {darkseagreen1}       {rgb}{0.76,1.00,0.76}
\definecolor {darkseagreen2}       {rgb}{0.71,0.93,0.71}
\definecolor {darkseagreen3}       {rgb}{0.61,0.80,0.61}
\definecolor {darkseagreen4}       {rgb}{0.41,0.55,0.41}
\definecolor {seagreen1}           {rgb}{0.33,1.00,0.62}
\definecolor {seagreen2}           {rgb}{0.31,0.93,0.58}
\definecolor {seagreen3}           {rgb}{0.26,0.80,0.50}
\definecolor {palegreen1}          {rgb}{0.60,1.00,0.60}
\definecolor {palegreen2}          {rgb}{0.56,0.93,0.56}
\definecolor {palegreen3}          {rgb}{0.49,0.80,0.49}
\definecolor {palegreen4}          {rgb}{0.33,0.55,0.33}
\definecolor {springgreen2}        {rgb}{0.00,0.93,0.46}
\definecolor {springgreen3}        {rgb}{0.00,0.80,0.40}
\definecolor {springgreen4}        {rgb}{0.00,0.55,0.27}
\definecolor {green2}              {rgb}{0.00,0.93,0.00}
\definecolor {green3}              {rgb}{0.00,0.80,0.00}
\definecolor {green4}              {rgb}{0.00,0.55,0.00}
\definecolor {chartreuse2}         {rgb}{0.46,0.93,0.00}
\definecolor {chartreuse3}         {rgb}{0.40,0.80,0.00}
\definecolor {chartreuse4}         {rgb}{0.27,0.55,0.00}
\definecolor {olivedrab1}          {rgb}{0.75,1.00,0.24}
\definecolor {olivedrab2}          {rgb}{0.70,0.93,0.23}
\definecolor {olivedrab4}          {rgb}{0.41,0.55,0.13}
\definecolor {darkolivegreen1}     {rgb}{0.79,1.00,0.44}
\definecolor {darkolivegreen2}     {rgb}{0.74,0.93,0.41}
\definecolor {darkolivegreen3}     {rgb}{0.64,0.80,0.35}
\definecolor {darkolivegreen4}     {rgb}{0.43,0.55,0.24}
\definecolor {khaki1}              {rgb}{1.00,0.96,0.56}
\definecolor {khaki2}              {rgb}{0.93,0.90,0.52}
\definecolor {khaki3}              {rgb}{0.80,0.78,0.45}
\definecolor {khaki4}              {rgb}{0.55,0.53,0.31}
\definecolor {lightgoldenrod1}     {rgb}{1.00,0.93,0.55}
\definecolor {lightgoldenrod2}     {rgb}{0.93,0.86,0.51}
\definecolor {lightgoldenrod3}     {rgb}{0.80,0.75,0.44}
\definecolor {lightgoldenrod4}     {rgb}{0.55,0.51,0.30}
\definecolor {lightyellow2}        {rgb}{0.93,0.93,0.82}
\definecolor {lightyellow3}        {rgb}{0.80,0.80,0.71}
\definecolor {lightyellow4}        {rgb}{0.55,0.55,0.48}
\definecolor {yellow2}             {rgb}{0.93,0.93,0.00}
\definecolor {yellow3}             {rgb}{0.80,0.80,0.00}
\definecolor {yellow4}             {rgb}{0.55,0.55,0.00}
\definecolor {gold2}               {rgb}{0.93,0.79,0.00}
\definecolor {gold3}               {rgb}{0.80,0.68,0.00}
\definecolor {gold4}               {rgb}{0.55,0.46,0.00}
\definecolor {goldenrod1}          {rgb}{1.00,0.76,0.15}
\definecolor {goldenrod2}          {rgb}{0.93,0.71,0.13}
\definecolor {goldenrod3}          {rgb}{0.80,0.61,0.11}
\definecolor {goldenrod4}          {rgb}{0.55,0.41,0.08}
\definecolor {darkgoldenrod1}      {rgb}{1.00,0.73,0.06}
\definecolor {darkgoldenrod2}      {rgb}{0.93,0.68,0.05}
\definecolor {darkgoldenrod3}      {rgb}{0.80,0.58,0.05}
\definecolor {darkgoldenrod4}      {rgb}{0.55,0.40,0.03}
\definecolor {rosybrown1}          {rgb}{1.00,0.76,0.76}
\definecolor {rosybrown2}          {rgb}{0.93,0.71,0.71}
\definecolor {rosybrown3}          {rgb}{0.80,0.61,0.61}
\definecolor {rosybrown4}          {rgb}{0.55,0.41,0.41}
\definecolor {indianred1}          {rgb}{1.00,0.42,0.42}
\definecolor {indianred2}          {rgb}{0.93,0.39,0.39}
\definecolor {indianred3}          {rgb}{0.80,0.33,0.33}
\definecolor {indianred4}          {rgb}{0.55,0.23,0.23}
\definecolor {sienna1}             {rgb}{1.00,0.51,0.28}
\definecolor {sienna2}             {rgb}{0.93,0.47,0.26}
\definecolor {sienna3}             {rgb}{0.80,0.41,0.22}
\definecolor {sienna4}             {rgb}{0.55,0.28,0.15}
\definecolor {burlywood1}          {rgb}{1.00,0.83,0.61}
\definecolor {burlywood2}          {rgb}{0.93,0.77,0.57}
\definecolor {burlywood3}          {rgb}{0.80,0.67,0.49}
\definecolor {burlywood4}          {rgb}{0.55,0.45,0.33}
\definecolor {wheat1}              {rgb}{1.00,0.91,0.73}
\definecolor {wheat2}              {rgb}{0.93,0.85,0.68}
\definecolor {wheat3}              {rgb}{0.80,0.73,0.59}
\definecolor {wheat4}              {rgb}{0.55,0.49,0.40}
\definecolor {tan1}                {rgb}{1.00,0.65,0.31}
\definecolor {tan2}                {rgb}{0.93,0.60,0.29}
\definecolor {tan4}                {rgb}{0.55,0.35,0.17}
\definecolor {chocolate1}          {rgb}{1.00,0.50,0.14}
\definecolor {chocolate2}          {rgb}{0.93,0.46,0.13}
\definecolor {chocolate3}          {rgb}{0.80,0.40,0.11}
\definecolor {firebrick1}          {rgb}{1.00,0.19,0.19}
\definecolor {firebrick2}          {rgb}{0.93,0.17,0.17}
\definecolor {firebrick3}          {rgb}{0.80,0.15,0.15}
\definecolor {firebrick4}          {rgb}{0.55,0.10,0.10}
\definecolor {brown1}              {rgb}{1.00,0.25,0.25}
\definecolor {brown2}              {rgb}{0.93,0.23,0.23}
\definecolor {brown3}              {rgb}{0.80,0.20,0.20}
\definecolor {brown4}              {rgb}{0.55,0.14,0.14}
\definecolor {salmon1}             {rgb}{1.00,0.55,0.41}
\definecolor {salmon2}             {rgb}{0.93,0.51,0.38}
\definecolor {salmon3}             {rgb}{0.80,0.44,0.33}
\definecolor {salmon4}             {rgb}{0.55,0.30,0.22}
\definecolor {lightsalmon2}        {rgb}{0.93,0.58,0.45}
\definecolor {lightsalmon3}        {rgb}{0.80,0.51,0.38}
\definecolor {lightsalmon4}        {rgb}{0.55,0.34,0.26}
\definecolor {orange2}             {rgb}{0.93,0.60,0.00}
\definecolor {orange3}             {rgb}{0.80,0.52,0.00}
\definecolor {orange4}             {rgb}{0.55,0.35,0.00}
\definecolor {darkorange1}         {rgb}{1.00,0.50,0.00}
\definecolor {darkorange2}         {rgb}{0.93,0.46,0.00}
\definecolor {darkorange3}         {rgb}{0.80,0.40,0.00}
\definecolor {darkorange4}         {rgb}{0.55,0.27,0.00}
\definecolor {coral1}              {rgb}{1.00,0.45,0.34}
\definecolor {coral2}              {rgb}{0.93,0.42,0.31}
\definecolor {coral3}              {rgb}{0.80,0.36,0.27}
\definecolor {coral4}              {rgb}{0.55,0.24,0.18}
\definecolor {tomato2}             {rgb}{0.93,0.36,0.26}
\definecolor {tomato3}             {rgb}{0.80,0.31,0.22}
\definecolor {tomato4}             {rgb}{0.55,0.21,0.15}
\definecolor {orangered2}          {rgb}{0.93,0.25,0.00}
\definecolor {orangered3}          {rgb}{0.80,0.22,0.00}
\definecolor {orangered4}          {rgb}{0.55,0.15,0.00}
\definecolor {red2}                {rgb}{0.93,0.00,0.00}
\definecolor {red3}                {rgb}{0.80,0.00,0.00}
\definecolor {red4}                {rgb}{0.55,0.00,0.00}
\definecolor {deeppink2}           {rgb}{0.93,0.07,0.54}
\definecolor {deeppink3}           {rgb}{0.80,0.06,0.46}
\definecolor {deeppink4}           {rgb}{0.55,0.04,0.31}
\definecolor {hotpink1}            {rgb}{1.00,0.43,0.71}
\definecolor {hotpink2}            {rgb}{0.93,0.42,0.65}
\definecolor {hotpink3}            {rgb}{0.80,0.38,0.56}
\definecolor {hotpink4}            {rgb}{0.55,0.23,0.38}
\definecolor {pink1}               {rgb}{1.00,0.71,0.77}
\definecolor {pink2}               {rgb}{0.93,0.66,0.72}
\definecolor {pink3}               {rgb}{0.80,0.57,0.62}
\definecolor {pink4}               {rgb}{0.55,0.39,0.42}
\definecolor {lightpink1}          {rgb}{1.00,0.68,0.73}
\definecolor {lightpink2}          {rgb}{0.93,0.64,0.68}
\definecolor {lightpink3}          {rgb}{0.80,0.55,0.58}
\definecolor {lightpink4}          {rgb}{0.55,0.37,0.40}
\definecolor {palevioletred1}      {rgb}{1.00,0.51,0.67}
\definecolor {palevioletred2}      {rgb}{0.93,0.47,0.62}
\definecolor {palevioletred3}      {rgb}{0.80,0.41,0.54}
\definecolor {palevioletred4}      {rgb}{0.55,0.28,0.36}
\definecolor {maroon1}             {rgb}{1.00,0.20,0.70}
\definecolor {maroon2}             {rgb}{0.93,0.19,0.65}
\definecolor {maroon3}             {rgb}{0.80,0.16,0.56}
\definecolor {maroon4}             {rgb}{0.55,0.11,0.38}
\definecolor {violetred1}          {rgb}{1.00,0.24,0.59}
\definecolor {violetred2}          {rgb}{0.93,0.23,0.55}
\definecolor {violetred3}          {rgb}{0.80,0.20,0.47}
\definecolor {violetred4}          {rgb}{0.55,0.13,0.32}
\definecolor {magenta2}            {rgb}{0.93,0.00,0.93}
\definecolor {magenta3}            {rgb}{0.80,0.00,0.80}
\definecolor {magenta4}            {rgb}{0.55,0.00,0.55}
\definecolor {orchid1}             {rgb}{1.00,0.51,0.98}
\definecolor {orchid2}             {rgb}{0.93,0.48,0.91}
\definecolor {orchid3}             {rgb}{0.80,0.41,0.79}
\definecolor {orchid4}             {rgb}{0.55,0.28,0.54}
\definecolor {plum1}               {rgb}{1.00,0.73,1.00}
\definecolor {plum2}               {rgb}{0.93,0.68,0.93}
\definecolor {plum3}               {rgb}{0.80,0.59,0.80}
\definecolor {plum4}               {rgb}{0.55,0.40,0.55}
\definecolor {mediumorchid1}       {rgb}{0.88,0.40,1.00}
\definecolor {mediumorchid2}       {rgb}{0.82,0.37,0.93}
\definecolor {mediumorchid3}       {rgb}{0.71,0.32,0.80}
\definecolor {mediumorchid4}       {rgb}{0.48,0.22,0.55}
\definecolor {darkorchid1}         {rgb}{0.75,0.24,1.00}
\definecolor {darkorchid2}         {rgb}{0.70,0.23,0.93}
\definecolor {darkorchid3}         {rgb}{0.60,0.20,0.80}
\definecolor {darkorchid4}         {rgb}{0.41,0.13,0.55}
\definecolor {purple1}             {rgb}{0.61,0.19,1.00}
\definecolor {purple2}             {rgb}{0.57,0.17,0.93}
\definecolor {purple3}             {rgb}{0.49,0.15,0.80}
\definecolor {purple4}             {rgb}{0.33,0.10,0.55}
\definecolor {mediumpurple1}       {rgb}{0.67,0.51,1.00}
\definecolor {mediumpurple2}       {rgb}{0.62,0.47,0.93}
\definecolor {mediumpurple3}       {rgb}{0.54,0.41,0.80}
\definecolor {mediumpurple4}       {rgb}{0.36,0.28,0.55}
\definecolor {thistle1}            {rgb}{1.00,0.88,1.00}
\definecolor {thistle2}            {rgb}{0.93,0.82,0.93}
\definecolor {thistle3}            {rgb}{0.80,0.71,0.80}
\definecolor {thistle4}            {rgb}{0.55,0.48,0.55}
\definecolor {gray1}               {rgb}{0.01,0.01,0.01}
\definecolor {gray2}               {rgb}{0.02,0.02,0.02}
\definecolor {gray3}               {rgb}{0.03,0.03,0.03}
\definecolor {gray4}               {rgb}{0.04,0.04,0.04}
\definecolor {gray5}               {rgb}{0.05,0.05,0.05}
\definecolor {gray6}               {rgb}{0.06,0.06,0.06}
\definecolor {gray7}               {rgb}{0.07,0.07,0.07}
\definecolor {gray8}               {rgb}{0.08,0.08,0.08}
\definecolor {gray9}               {rgb}{0.09,0.09,0.09}
\definecolor {gray10}              {rgb}{0.10,0.10,0.10}
\definecolor {gray11}              {rgb}{0.11,0.11,0.11}
\definecolor {gray12}              {rgb}{0.12,0.12,0.12}
\definecolor {gray13}              {rgb}{0.13,0.13,0.13}
\definecolor {gray14}              {rgb}{0.14,0.14,0.14}
\definecolor {gray15}              {rgb}{0.15,0.15,0.15}
\definecolor {gray16}              {rgb}{0.16,0.16,0.16}
\definecolor {gray17}              {rgb}{0.17,0.17,0.17}
\definecolor {gray18}              {rgb}{0.18,0.18,0.18}
\definecolor {gray19}              {rgb}{0.19,0.19,0.19}
\definecolor {gray20}              {rgb}{0.20,0.20,0.20}
\definecolor {gray21}              {rgb}{0.21,0.21,0.21}
\definecolor {gray22}              {rgb}{0.22,0.22,0.22}
\definecolor {gray23}              {rgb}{0.23,0.23,0.23}
\definecolor {gray24}              {rgb}{0.24,0.24,0.24}
\definecolor {gray25}              {rgb}{0.25,0.25,0.25}
\definecolor {gray26}              {rgb}{0.26,0.26,0.26}
\definecolor {gray27}              {rgb}{0.27,0.27,0.27}
\definecolor {gray28}              {rgb}{0.28,0.28,0.28}
\definecolor {gray29}              {rgb}{0.29,0.29,0.29}
\definecolor {gray30}              {rgb}{0.30,0.30,0.30}
\definecolor {gray31}              {rgb}{0.31,0.31,0.31}
\definecolor {gray32}              {rgb}{0.32,0.32,0.32}
\definecolor {gray33}              {rgb}{0.33,0.33,0.33}
\definecolor {gray34}              {rgb}{0.34,0.34,0.34}
\definecolor {gray35}              {rgb}{0.35,0.35,0.35}
\definecolor {gray36}              {rgb}{0.36,0.36,0.36}
\definecolor {gray37}              {rgb}{0.37,0.37,0.37}
\definecolor {gray38}              {rgb}{0.38,0.38,0.38}
\definecolor {gray39}              {rgb}{0.39,0.39,0.39}
\definecolor {gray40}              {rgb}{0.40,0.40,0.40}
\definecolor {gray42}              {rgb}{0.42,0.42,0.42}
\definecolor {gray43}              {rgb}{0.43,0.43,0.43}
\definecolor {gray44}              {rgb}{0.44,0.44,0.44}
\definecolor {gray45}              {rgb}{0.45,0.45,0.45}
\definecolor {gray46}              {rgb}{0.46,0.46,0.46}
\definecolor {gray47}              {rgb}{0.47,0.47,0.47}
\definecolor {gray48}              {rgb}{0.48,0.48,0.48}
\definecolor {gray49}              {rgb}{0.49,0.49,0.49}
\definecolor {gray50}              {rgb}{0.50,0.50,0.50}
\definecolor {gray51}              {rgb}{0.51,0.51,0.51}
\definecolor {gray52}              {rgb}{0.52,0.52,0.52}
\definecolor {gray53}              {rgb}{0.53,0.53,0.53}
\definecolor {gray54}              {rgb}{0.54,0.54,0.54}
\definecolor {gray55}              {rgb}{0.55,0.55,0.55}
\definecolor {gray56}              {rgb}{0.56,0.56,0.56}
\definecolor {gray57}              {rgb}{0.57,0.57,0.57}
\definecolor {gray58}              {rgb}{0.58,0.58,0.58}
\definecolor {gray59}              {rgb}{0.59,0.59,0.59}
\definecolor {gray60}              {rgb}{0.60,0.60,0.60}
\definecolor {gray61}              {rgb}{0.61,0.61,0.61}
\definecolor {gray62}              {rgb}{0.62,0.62,0.62}
\definecolor {gray63}              {rgb}{0.63,0.63,0.63}
\definecolor {gray64}              {rgb}{0.64,0.64,0.64}
\definecolor {gray65}              {rgb}{0.65,0.65,0.65}
\definecolor {gray66}              {rgb}{0.66,0.66,0.66}
\definecolor {gray67}              {rgb}{0.67,0.67,0.67}
\definecolor {gray68}              {rgb}{0.68,0.68,0.68}
\definecolor {gray69}              {rgb}{0.69,0.69,0.69}
\definecolor {gray70}              {rgb}{0.70,0.70,0.70}
\definecolor {gray71}              {rgb}{0.71,0.71,0.71}
\definecolor {gray72}              {rgb}{0.72,0.72,0.72}
\definecolor {gray73}              {rgb}{0.73,0.73,0.73}
\definecolor {gray74}              {rgb}{0.74,0.74,0.74}
\definecolor {gray75}              {rgb}{0.75,0.75,0.75}
\definecolor {gray76}              {rgb}{0.76,0.76,0.76}
\definecolor {gray77}              {rgb}{0.77,0.77,0.77}
\definecolor {gray78}              {rgb}{0.78,0.78,0.78}
\definecolor {gray79}              {rgb}{0.79,0.79,0.79}
\definecolor {gray80}              {rgb}{0.80,0.80,0.80}
\definecolor {gray81}              {rgb}{0.81,0.81,0.81}
\definecolor {gray82}              {rgb}{0.82,0.82,0.82}
\definecolor {gray83}              {rgb}{0.83,0.83,0.83}
\definecolor {gray84}              {rgb}{0.84,0.84,0.84}
\definecolor {gray85}              {rgb}{0.85,0.85,0.85}
\definecolor {gray86}              {rgb}{0.86,0.86,0.86}
\definecolor {gray87}              {rgb}{0.87,0.87,0.87}
\definecolor {gray88}              {rgb}{0.88,0.88,0.88}
\definecolor {gray89}              {rgb}{0.89,0.89,0.89}
\definecolor {gray90}              {rgb}{0.90,0.90,0.90}
\definecolor {gray91}              {rgb}{0.91,0.91,0.91}
\definecolor {gray92}              {rgb}{0.92,0.92,0.92}
\definecolor {gray93}              {rgb}{0.93,0.93,0.93}
\definecolor {gray94}              {rgb}{0.94,0.94,0.94}
\definecolor {gray95}              {rgb}{0.95,0.95,0.95}
\definecolor {gray97}              {rgb}{0.97,0.97,0.97}
\definecolor {gray98}              {rgb}{0.98,0.98,0.98}
\definecolor {gray99}              {rgb}{0.99,0.99,0.99}
\definecolor {darkgrey}            {rgb}{0.66,0.66,0.66}

\renewcommand{\algorithmiccomment}[1]{\ \ \ \ \ // #1}

\newcommand{\TODO}[1]{{}}

\newcommand{\ignore}[1]{}
\newcommand{\todo}[1] {}

\newcommand{\RSTODO}[1]{{\bf \textcolor{darkgreen}{{\fbox{RS TODO:} #1}}}}
\renewcommand{\RSTODO}[1]{}

\newenvironment{rs}{\color{darkgreen}}{\normalcolor}

 \newcommand{\ignoreinshort}[1]{}
 \newcommand{\ignoreinlong}[1]{{#1}}
\def\makenewenumerate#1#2{%
\newcounter{cnt#1}
\newenvironment{#1}%
{\begin{list}{\makebox[0pt][r]{#2}}%
{\setlength{\itemsep}{0pt}% 
 \setlength{\parsep}{.2em}%
 \setlength{\leftmargin}{1.5em}%
 \setlength{\labelwidth}{.4em}%
 \usecounter{cnt#1}}}
{\end{list}}}

\makenewenumerate{myenumerate}{\arabic{cntmyenumerate}.}

\makenewenumerate{renumerate}{\rm(\roman{cntrenumerate})}
\makenewenumerate{renumerateprime}{\rm(\roman{cntrenumerateprime}$'$)}
\makenewenumerate{renumeratesecond}{\rm(\roman{cntrenumeratesecond}$''$)}

\makenewenumerate{aenumerate}{({\it\alph{cntaenumerate}})}
\makenewenumerate{aenumerateprime}{({\it\alph{cntaenumerateprime}$'$})}
\makenewenumerate{aenumeratesecond}{({\it\alph{cntaenumeratesecond}$''$})}

\def\newplaintheorem#1#2{%
\newtheorem{#1plain}{#2}[section]%
\newenvironment{#1}{\begin{#1plain}\rm }{\end{#1plain}}}

\newplaintheorem{notation}{Notation}
\newplaintheorem{cexample}{Counter-Example}

\newcommand{\fakesubsubsection}[1]{\smallskip\noindent {\bf #1.}}

\newcommand{\sref}[1]{\S{}\ref{#1}}
\newcommand{\noi}{\noindent}

\newcommand{\pair}[2]{\ensuremath{\langle{#1},{#2}\rangle}\xspace}

\newcommand{\tuple}[1]{\ensuremath{\langle{#1}\rangle}\xspace}
\newcommand{\set}[1]{\ensuremath{\{{#1}\}}\xspace}
\newcommand{\imp}{\ensuremath{\rightarrow}\xspace}

\newcommand{\defas}{\ensuremath{\stackrel{\text{\tiny def}}{=}}\xspace}

\newcommand\calc{\ensuremath{\mathcal{C}}\xspace}

\newcommand\calm{\ensuremath{\mathcal{M}}\xspace}

\newcommand{\omt}{\ensuremath{\text{OMT}}\xspace}

\newcommand{\omlarat}{\ensuremath{\text{OMT}(\larat)}\xspace}
\newcommand{\omlaint}{\ensuremath{\text{OMT}(\laint)}\xspace}
\newcommand{\omla}{\ensuremath{\text{OMT}(\la)}\xspace}

\newcommand{\omlaratplus}{\ensuremath{\text{OMT}(\larat\cup\T)}\xspace}
\newcommand{\omlaplus}{\ensuremath{\text{OMT}(\la\cup\T)}\xspace}

\newcommand{\smtlaratplus}{\smttt{\larat\cup\T}\xspace}

\newcommand{\lbgen}[2]{\ensuremath{{\sf lb}^{#1}_{#2}}\xspace}
\newcommand{\ubgen}[2]{\ensuremath{{\sf ub}^{#1}_{#2}}\xspace}
\newcommand{\lb}{\lbgen{}{}}
\newcommand{\ub}{\ubgen{}{}}

\newcommand{\pivot}{\ensuremath{\mathsf{pivot}}\xspace}
\newcommand{\trueval}{{\ensuremath{\mathsf{true}}}}

\newcommand{\cost}{\ensuremath{{cost}}\xspace}
\newcommand{\mincost}{\ensuremath{{mincost}}\xspace}

\newcommand{\C}{\ensuremath{\mathcal{C}}\xspace}

\newcommand\mysout{\bgroup \markoverwith{{-}}\ULon}
\newcommand\nosout{\bgroup \markoverwith{{ }}\ULon}
\definecolor{mygray}{rgb}{0.90,0.90,0.90}
\definecolor{mywhite}{rgb}{1.00,1.00,1.00}

\newcommand{\currlb}{\ensuremath{\mathsf{l}}\xspace}
\newcommand{\currub}{\ensuremath{\mathsf{u}}\xspace}
\newcommand{\range}{\ensuremath{[\lb,\ub[}\xspace}
\newcommand{\currrange}{\ensuremath{[\currlb,\currub[}\xspace}
\newcommand{\lpivotrange}{\ensuremath{[\currlb,\pivot[}\xspace}
\newcommand{\rpivotrange}{\ensuremath{[\pivot,\currub[}\xspace}

\newcommand{\dopivoting}{\ensuremath{{\sf BinSearchMode()}}\xspace}

\newcommand{\ubliti}[1]{\ensuremath{(\cost < #1)}\xspace}
\newcommand{\lbliti}[1]{\ensuremath{\neg(\cost < #1)}\xspace}
\newcommand{\pivotatom}{\ensuremath{\mathsf{PIV}}\xspace}

\newcommand{\minvalue}{\ensuremath{\mathsf{min}}\xspace}
\newcommand{\maxvalue}{\ensuremath{\mathsf{max}}\xspace}

\renewcommand{\mincost}{\ensuremath{\mathsf{min}_\cost}\xspace}

\newcommand{\optimathsat}{\textsc{OptiMathSAT}\xspace}

\newcommand{\symba}{\textsc{Symba}\xspace}

\newcommand{\satres}{\textsc{sat}\xspace}
\newcommand{\unsatres}{\textsc{unsat}\xspace}

\newcommand{\vi}{\ensuremath{\varphi}\xspace}
\newcommand{\vip}{\ensuremath{\varphi^p}\xspace}
\newcommand{\mup}{\ensuremath{\mu^p}\xspace}

\newcommand{\etap}{\ensuremath{\eta^p}\xspace}

\newcommand{\T}{\ensuremath{\mathcal{T}}\xspace}

\newcommand{\smt}{SMT\xspace}
\newcommand{\smtt}{\ensuremath{\text{SMT}(\T)}\xspace}
\newcommand{\smttt}[1]{\ensuremath{\text{SMT}(#1)}\xspace}

\newcommand{\euf}{\ensuremath{\mathcal{EUF}}\xspace}

\newcommand{\la}{\ensuremath{\mathcal{LA}}\xspace}
\newcommand{\larat}{\ensuremath{\mathcal{LA}(\mathbb{Q})}\xspace}
\newcommand{\laint}{\ensuremath{\mathcal{LA}(\mathbb{Z})}\xspace}

\renewcommand{\la}{\ensuremath{\mathcal{LA}}\xspace}
\renewcommand{\larat}{\ensuremath{\mathcal{LRA}}\xspace}
\renewcommand{\laint}{\ensuremath{\mathcal{LIA}}\xspace}

\newcommand{\nlarat}{\ensuremath{\mathcal{NLA}(\mathbb{R})}\xspace}
\newcommand{\nlaint}{\ensuremath{\mathcal{NLA}(\mathbb{Z})}\xspace}
\newcommand{\bv}{\ensuremath{\mathcal{BV}}\xspace}
\newcommand{\mem}{\ensuremath{\mathcal{AR}}\xspace}

\newcommand{\pmodels}{\models_p}

\newcommand{\TsolverGen}[1]{\ensuremath{{#1}\textit{-solver}}\xspace}
\newcommand{\TsolversGen}[1]{\ensuremath{{#1}\textit{-solvers}}\xspace}
\newcommand{\Tsolver}{\TsolverGen{\T}}
\newcommand{\Tsolvers}{\TsolversGen{\T}}

\newcommand{\Tlemmas}{\T-lemmas\xspace}

\newcommand{\laratsolver}{\larat-\ensuremath{\mathsf{Solver}}}

\newcommand{\laintsolver}{\laint-\ensuremath{\mathsf{Solver}}}

\newcommand{\lasolver}{\la-\ensuremath{\mathsf{Solver}}}

\newcommand{\mathsat}{\textsc{MathSAT}\xspace}

\newcommand{\yices}{\textsc{Yices}\xspace}

\newcommand{\zthree}{\textsc{Z3}\xspace}

\newcommand{\mathsatfive}{\textsc{MathSAT5}\xspace}

\newcommand{\optzt}{\textsc{opt-z3}\xspace}

\newcommand{\PTTODO}[1]{}

\renewcommand{\RSTODO}[1]{{\bf \textcolor{blue}{{\fbox{RS TODO:} #1}}}}
\renewcommand{\PTTODO}[1]{{\bf \textcolor{darkgreen}{{\fbox{PT TODO:} #1}}}}

\newcommand{\currentcosts}{\ensuremath{\calc^*}}

\newcommand{\localbounds}[1]{}
\newcommand{\globalbounds}[1]{}

\newcommand{\laintminimize}{\laint-\ensuremath{\mathsf{Minimize}}\xspace}
\newcommand{\laratminimize}{\larat-\ensuremath{\mathsf{Minimize}}\xspace}
\newcommand{\laminimize}{\la-\ensuremath{\mathsf{Minimize}}\xspace}

\newcommand{\uncutifnecessary}[1]{}

\renewcommand{\nlarat}{\ensuremath{\mathcal{NLRA}}\xspace}
\renewcommand{\nlaint}{\ensuremath{\mathcal{NLIA}}\xspace}

\renewcommand{\la}{\ensuremath{\mathcal{LRIA}}\xspace}
\renewcommand{\lasolver}{\la-\ensuremath{\mathsf{Solver}}\xspace}

\renewcommand{\laminimize}{\la-\ensuremath{\mathsf{Minimize}}\xspace}
\renewcommand{\omla}{\ensuremath{\text{OMT}(\la)}\xspace}

\newcommand{\bclt}{\textsc{bclt}\xspace}
\newcommand{\sal}{\textsc{SAL}\xspace}

\newcommand{\nuz}{\textsc{$\nu Z$}\xspace}

\newcommand{\longversion}{true}

\ifthenelse{\equal{\longversion}{true}}
{%
  \renewcommand{\ignoreinshort}[1]{\textcolor{blue}{#1}}
  \renewcommand{\ignoreinshort}[1]{{#1}}
  \renewcommand{\ignoreinlong}[1]{}
}%
{%
 \renewcommand{\ignoreinshort}[1]{}
 \renewcommand{\ignoreinlong}[1]{{#1}}
}
\begin{document}

\pagestyle{plain}
\pagenumbering{roman}

\pagestyle{plain}
\pagenumbering{arabic}

\title{%
Pushing the Envelope of
Optimization Modulo Theories with 
Linear-Arithmetic Cost Functions
\thanks{This work is supported by %\ignoreinlong{SRC}
{Semiconductor Research Corporation} (SRC)
  under
GRC Research Project 2012-TJ-2266 WOLF.
{We thank Alberto Griggio for support with \mathsatfive code.}}
}

\author{
Roberto Sebastiani \and
Patrick Trentin
}

\institute{%
DISI, University of Trento, Italy%
}

\maketitle
\ignoreinshort{
\begin{center}
\noi
{\em NOTE\\
 This is an extended version of a paper published at TACAS 2015
\cite{st_tacas15}. 
\\Latest update: \today}
\end{center}
}
 \begin{abstract}
 In the last decade we have witnessed an impressive progress in the
expressiveness and efficiency of Satisfiability Modulo Theories (SMT)
solving techniques. This has brought previously-intractable problems
at the reach of state-of-the-art SMT solvers, in particular in the domain
of SW and HW verification.
Many SMT-encodable problems of interest, however, require also
 the capability of finding models that are {\em optimal} wrt. some
 cost functions. 
 In previous work, namely {\em Optimization Modulo Theory with Linear
 Rational Cost Functions -- \omlaratplus}, we have leveraged SMT
 solving to handle the {\em minimization} of cost functions on
 linear arithmetic over the rationals, by means of a combination of
 SMT and LP minimization techniques.

In this paper we push the envelope of our OMT approach along
three directions: 
first, we extend it to work with linear arithmetic on the
mixed integer/rational domain, 
  by means of a combination of SMT, LP and ILP minimization
  techniques;
second, we develop a {\em multi-objective}
version of OMT, so that to 
 handle many  cost functions simultaneously or lexicographically;  
third, we develop an {\em incremental} 
version of OMT, so that to 
exploit the incrementality of some OMT-encodable problems.  
An empirical evaluation performed on OMT-encoded verification
problems demonstrates the usefulness and
efficiency  of these extensions.

 \end{abstract}

\section{Introduction}
\label{sec:intro}
%\marg{SMT generalities}
In many contexts including automated reasoning (AR) and formal
verification (FV)
important {\em decision} problems are effectively encoded into and solved as
Satisfiability Modulo Theories (SMT) problems.
In the last decade efficient SMT
solvers have been developed, that combine the power of modern 
conflict-driven clause-learning (CDCL)  SAT solvers \ignoreinshort{\cite{MSLM09HBSAT}}
with the expressiveness of dedicated decision procedures (\Tsolvers)
for several first-order 
theories of practical interest like, e.g., those of linear arithmetic 
over the
rationals (\larat{}) or the integers (\laint{}) or their combination (\la), 
those of non-linear arithmetic over the reals (\nlarat) or the integers
(\nlaint),  of arrays (\mem),
of bit-vectors (\bv), and their combinations.
(See \cite{nieot-jacm-06,sebastiani07,BSST09HBSAT} for an overview.)
%\ignoreinlong{(See \cite{BSST09HBSAT} for an overview.)}
This has brought previously-intractable problems
at the reach of state-of-the-art SMT solvers, in particular in the domain
of software (SW) and hardware (HW) verification.
  
 %\marg{need for \omtt}%
 Many SMT-encodable problems of interest, however, may require also
 the capability of finding models that are {\em optimal} wrt. some
 cost function over  arithmetical variables.
(See e.g. \cite{st-ijcar12,li_popl14,st_tocl14} for a rich list of such applications.)
For instance, in
SMT-based {\em model checking with timed or hybrid systems}
{(e.g. \cite{acks_forte02,AudemardBCS05})}
you may want to find
executions which 
%minimize some parameter (e.g. elapsed time), or which
{optimize the value of some  parameter (e.g., a clock
timeout value, or the total elapsed time)} while fulfilling/violating some property (e.g.,
find the minimum time interval for a rail-crossing causing a safety
 violation).

%\marg{previous work}
Surprisingly, only few works 
extending SMT  
to deal with {\em optimization} problems have been presented in the literature
\cite{nieuwenhuis_sat06,cimattifgss10,st-ijcar12,dilligdma12,manoliosp13,cgss_sat13_maxsmt,st_tocl14,li_popl14,LarrazORR14,bjorner_scss14} 
--most of which handle problems which are different to that 
 addressed in this paper \cite{nieuwenhuis_sat06,cimattifgss10,dilligdma12,manoliosp13,cgss_sat13_maxsmt}\ignoreinshort{, 
see related work below}. 
\ignoreinlong{%}
(We refer the reader to the related work section of \cite{st_tocl14}
for a discussion on these approaches.)}

%\marg{\omlaratplus}
%In particular, %%%%%in \cite{st-ijcar12,st_tocl14} 
Sebastiani and Tomasi \cite{st-ijcar12,st_tocl14} 
% we~\footnote{
% By ``we''  we mean ``the \optimathsat team'', although its composition is
% in evolution.}
 presented two 
procedures for adding to
\smtlaratplus the functionality of finding
models minimizing some \larat cost variable 
--\T being some (possibly empty) stably-infinite theory
s.t. \T and \larat are signature-disjoint.
This problem is referred to as {\em Optimization Modulo Theories 
with linear  cost functions on the rationals}, \omlaratplus. 
(If \T is the empty theory, then we refer to it as \omlarat.)~%
\footnote{{Importantly, both MaxSMT (\cite{nieuwenhuis_sat06,cimattifgss10,cgss_sat13_maxsmt}) and SMT with pseudo-Boolean 
constraints and costs \cite{cimattifgss10} are straightforwardly encoded into OMT
\cite{st-ijcar12,st_tocl14}.}} 
These procedures  combine standard SMT
and LP minimization techniques:
the first, called {\em offline}, is much simpler to implement, since
it uses an incremental SMT solver as a black-box, whilst the second,
called {\em inline}, 
{embeds the search for optimum within the CDCL loop schema,
and as such 
it} is more sophisticate and efficient, but it
requires modifying the code of the SMT solver. 
% (This distinction is
%important, since the source code of many SMT solvers is not publicly
%available.)
In  \cite{st-ijcar12,st_tocl14} these procedures have been 
implemented on top of the  \mathsatfive  SMT solver
\cite{mathsat5_tacas13} into a tool called \optimathsat, and
 an extensive empirical evaluation is presented.

%\marg{\symba}
Li et al. \cite{li_popl14} extended 
  the \omlarat problem  %to ``multiple-objectives'',
by considering {\em contemporarily}
many cost functions  for the input formula \vi, namely
$\{\cost_1,...,\cost_k\}$, 
%\footnote{ 
so that the problem %\tuple{\vi,\{\cost_1,...,\cost_k\}}
  consists in enumerating $k$ independent models for \vi, each minimizing one
  specific $\cost_i$.~%
\ignoreinshort{\footnote{\ignoreinshort{More precisely, in \cite{li_popl14} the set of objectives
  {$k_1,k_2,...$} must be  {\em  maximized}, but the problem can be
  converted into a minimization problem by setting $\cost_i=-k_i$.
As in  \cite{li_popl14}, we remark also that this is {\em not} Pareto-optimality, where
a single model optimizing all objectives is searched.}}}
\ignoreinshort{(Intuitively, enumerating such models is in general more efficient than
solving one optimization problem at the time, because it allows for sharing the
SMT search steps among different cost objectives.)
}
In \cite{li_popl14} they presented a novel offline algorithm 
for \omlarat, and implemented it into the tool \symba.
Unlike with the procedures in  \cite{st-ijcar12,st_tocl14}, 
the algorithm described in \cite{li_popl14} does not use a LP minimization
 procedure: rather, a sequence of blackbox calls to an underlying SMT
solver (\zthree) allows for finding progressively-better solutions along some
objective direction, either forcing discrete jumps to some bounds induced by the
inequalities in the problem, or proving such objective is unbounded. 
\symba is used as backend engine of the SW model checker {\sc UFO}.~%
\footnote{\url{https://bitbucket.org/arieg/ufo/}}
An empirical evaluation on problems derived from SW verification shows
the usefulness of this multiple-cost approach.

Larraz et al. \cite{LarrazORR14} present
incomplete SMT(\nlaint) and MaxSMT(\nlaint) procedures, which use
an  \omlaint tool as an internal component. The latter
procedure, called \bclt,
 is described neither in \cite{LarrazORR14} nor in any 
previous publication;
however, it has been kindly made available to us  by
their authors upon request, together with a link to the master student's thesis 
describing it.~\footnote{\url{http://upcommons.upc.edu/pfc/handle/2099.1/14204?locale=en}.} 
 
%\newpage
%TODO{Includere riferimento a: 
%\cite{bjorner_scss14}
%\cite{nieuwenhuis14}
%}  
%\begin{remark}
{Finally, we have been informed by a reviewer of an invited presentation
given by  Bj{\o}rner and Phan two months after the submission of
%during the submission period of 
this paper  \cite{bjorner_scss14}, describing %who presented
general algorithms for optimization in SMT, including MaxSMT,
incremental, multi-objective 
 and lexicographic OMT, Pareto-optimality, which are implemented
into the tool \nuz{} on top of \zthree. %\footnote{\url{http://rise4fun.com/z3opt}.}
Remarkably, \cite{bjorner_scss14} presents specialized procedures for
MaxSMT, and enriches the offline OMT schema of
\cite{st-ijcar12,st_tocl14} with specialized algorithms 
for unbound-solution detection and for bound-tightening.
\ignore{%
(Thus, we have added  also a last-minute  empirical comparison with  \nuz;
%the tool in \cite{bjorner_scss14};
%
%TODO{REWRITE WHAT FOLLOWS: eye-catching difference...\\}
%Due to lack of time and space, 
more details are in an extended version
of this paper.~\footnote{Available at \url{http://optimathsat.disi.unitn.it}.})
}
\ignore{%
(Thus, we have added also a brief empirical comparison with 
the tool in \cite{bjorner_scss14}.)
}
}
%
%\end{remark}

We are not aware of any other OMT tool currently available.

%\marg{incrementality issue}
{We remark a few facts about the OMT tools in \cite{st-ijcar12,st_tocl14,li_popl14,LarrazORR14}.
First, none of them} has an 
{\em incremental} interface, allowing for pushing and popping subformulas 
(including definitions of novel cost functions) so that to reuse
previous search from one call to the other; in a FV
context this limitation is relevant, because often SMT backends 
are called incrementally (e.g., in the previously-mentioned example 
of SMT-based bounded model checking of timed\&hybrid systems).
%
%\marg{features}
Second, none of the above tools supports mixed integer/real
optimization, \omla.
Third,
none of the above tools supports {\em both} multi-objective
optimization and integer optimization. 
Finally, neither \symba nor \bclt currently handle combined theories. 

\smallskip
%\marg{Goal of the paper}
In this paper we push the envelope of the \omlaratplus approach 
of \cite{st-ijcar12,st_tocl14} along three directions: 
%
% \begin{renumerate}
% \item
(i) 
we extend it to work also with linear arithmetic on the
%{integer} and on 
mixed integer/rational domain, \omlaplus,
  by means of a combination of SMT, LP and ILP minimization
  techniques;
%\item 
(ii)
we develop a {\em multi-objective}
version of OMT, so that to 
 handle many  cost functions simultaneously  or lexicographically;  
(iii) we develop an {\em incremental} 
version of OMT, so that to 
exploit the incrementality of some OMT-encodable problems.  
%\end{renumerate}
% TODO{Evidenziare che nessuna delle procedure di cui sopra ha tutte
%   queste cose insieme.}
%
We have implement these novel functionalities in \optimathsat.
An empirical evaluation performed on OMT-encoded formal verification 
problems demonstrates the usefulness and
efficiency  of these extensions.

% \ignore{%
%(Thus, we have added  also a last-minute  empirical comparison with  \nuz;
%the tool in \cite{bjorner_scss14};
%
%TODO{REWRITE WHAT FOLLOWS: eye-catching difference...\\}
%Due to lack of time and space, 
\ignoreinlong{Some more details can be found in an extended version
of this paper.~\footnote{Available at \url{http://optimathsat.disi.unitn.it}.}
}

% : an {\em offline} and an {\em
% inline} procedure.  
% 

\paragraph{Content.} 
The paper is organized as follows: 
in \sref{sec:background} we provide the necessary background knowledge
on SMT and OMT;
in \sref{sec:pushing} we introduce and discuss the above-mentioned novel extensions of OMT; 
in \sref{sec:eval}  we perform an empirical evaluation of such procedures.
%in \sref{} 

\ignoreinshort{%
\subsubsection{Other Related Work}
\label{sec:related}
%
%\RSTODO{SFRONDARE QUANTO SEGUE?\\}
%
%\marg{Optimiz. in SMT}%
The idea of optimization in SMT was first introduced by Nieuwenhuis \&
Oliveras \cite{nieuwenhuis_sat06}, who presented an abstract logical 
framework of ``SMT with progressively stronger theories''
\ignoreinshort{(e.g., where the theory is progressively strengthened by every
new approximation of the minimum cost),}
and present implementations for MaxSMT based on this framework.
%
% Although in principle the logical formalism in \cite{nieuwenhuis_sat06} 
% works is not restricted to a particular king of cost functions,
%
Cimatti et al. \cite{cimattifgss10} introduced the notion of
 ``Theory of Costs'' \C to handle PB cost functions and constraints by
an ad-hoc and independent ``\C-solver'' in the standard lazy SMT
schema, and implemented a variant of MathSAT tool able to handle SMT
with PB constraints and to minimize PB cost functions. 
\ignore{%%%%%%%%% OTHERS
The SMT solvers \yices \cite{yices_sd} and \zthree \cite{z3_sd}
 also provide support for MaxSMT,
although there is no publicly-available document describing the
procedures used there. 
Ans{\'o}tegui et al. \cite{AnsoteguiBPSV11} describe the evaluation of
an implementation of a MaxSMT procedure based on \yices, although
%%RS added:
this implementation is not publicly available.
}
Cimatti et al. \cite{cgss_sat13_maxsmt}
presented a  ``modular'' approach for
MaxSMT, combining a lazy \smt{} solver with a
MaxSAT solver, which can be used as blackboxes.
We recall that SMT with PB functions and MaxSMT can be encoded
into each other, and that both are strictly less general than 
the \omlaratplus problems (see \cite{st-ijcar12,st_tocl14}).

% We recall that the problem addressed in this paper is strictly more general than
% both MaxSMT and SMT with PB functions (\sref{sec:optsmt}).

% \sttodo{aggiungere il confronto con  MaxSMT descritto in
% \cite{cgss_sat13_maxsmt}.
% reference to McMillan's work \cite{dilligdma12} and to ILP modulo
% theories \cite{manoliosp13}. Anche lavoro di Gurfinkel \& c.
% \cite{li_popl14}}

%\RSTODO{METTERE A POSTO DA QUI A MILP\\}

\smallskip
Two other forms of optimization in SMT, which are quite different from
the one presented in our work,  have been proposed in the
literature. 
Dillig et al. \cite{dilligdma12} addressed the problem of finding 
\emph{partial} models for quantified first-order
formulas modulo theories, which minimize the number of free
variables which are assigned a value from the domain.%
\ignoreinshort{%
Quoting an example from \cite{dilligdma12}, 
given the  formula {$\vi\defas(x+y+w>0)\vee(x+y+z+w<5)$},
the partial
  assignment $\{z=0\}$ satisfies
\vi because every total assignment extending it 
satisfies \vi and is minimum because there is no assignment satisfying \vi
which assigns less then one variable.
}
They proposed a general procedure addressing the problem for every
theory \T admitting quantifier elimination, and implemented a version 
for \laint and \euf into the {\sc Mistral} tool.
Manolios and Papavasileiou \cite{manoliosp13} proposed the 
``ILP Modulo Theories'' framework as an alternative to SAT Modulo
Theories, which allows for combining 
Integer Linear Programming with decision procedures for 
signature-disjoint stably-infinite theories \T;~%
they presented a general algorithm by integrating the Branch\&Cut 
ILP method with \T-specific decision procedures, and implemented it
into the {\sc Inez} tool.
\ignoreinshort{%
%\footnote{%
% According to Definition 2 in  \cite{manoliosp13}, 
% this approach 
%  cannot combine ILP with
%   \larat, since \laint and \larat are not signature-disjoint. 
%
%\\
 Notice that the approach of \cite{manoliosp13}
  cannot combine ILP with
   \larat, since \laint and \larat are not signature-disjoint. 
 (See Definition 2 in  \cite{manoliosp13}.)
Also, the objective
function is defined on the Integer domain. 
%aaaa}
%
}
We understand that %We understand that 
%to the best of our knowledge %and understanding, 
neither of the above-mentioned works can handle the
problem addressed in this paper, and vice versa.~%
(See \cite{st_tocl14} for a discussion on this topic.)

}

\section{Background}
\label{sec:background}
%
%\todo{move to background}
%
%\RSTODO{SFRONDARE QUANTO NON NECESSARIO\\}

\ignoreinshort{In this section we provide the necessary background on SMT and OMT.}

\subsection{Satisfiability Modulo Theories}
\label{sec:background_smt}
\noindent
{We assume a basic background knowledge on first-order logic and on
CDCL SAT solving\ignoreinshort{\ \cite{MSLM09HBSAT}}.
We consider some first-order theory \T, and we restrict our interest
to {\em ground} formulas/literals/atoms in the language of \T
(\T-formulas/literals/atoms hereafter).}~%
% NOTE (p.t.): dropped as reviewer 1 suggested
%\footnote{%
%Although we refer to quantifier-free formulas, as it is frequent
% practice in SAT and SMT, with a little abuse of terminology we often
% call ``Boolean variables'' the propositional atoms  and we call
% ``variables'' the
% Skolem constants $x_i$ in \larat-atoms like, e.g., ``$3x_1-2x_2+x_3\le 3$''.
% %
%}  
%

A {\em theory solver for \T}, \Tsolver, is a procedure able to
decide the \T-satisfiability of a conjunction/set $\mu$ of
\T-literals.
If $\mu$ is \T-unsatisfiable, then \Tsolver returns
\unsatres and a set/conjunction $\eta$ of \T-literals in $\mu$ which was
found \T-unsatisfiable;  $\eta$ is called a
\emph{\T-conflict set}, and $\neg\eta$ a \emph{\T-conflict clause}.~%
If $\mu$ is \T-satisfiable, then \Tsolver returns \satres; it may also
be able to return some unassigned \T-literal $l \not\in \mu$
 from a set of all available \T-literals,
% \footnote{Taken from a set of all the available \T-literals; when combined with a SAT solver, such set would be the set of all
% the \T-literals occurring in the input formula to solve.}
s.t. $\{l_1,...,l_n\}\models_{\T} l$, where
$\{l_1,...,l_n\}\subseteq\mu$.  We call this process
{\em \T-deduction} and $(\bigvee_{i=1}^n \neg l_i \vee l)$ a {\em
  \T-deduction clause}.
Notice that \T-conflict and \T-deduction clauses are valid in
\T. We call them {\em \Tlemmas}.

Given a \T-formula $\vi$, 
the formula \vip obtained by rewriting each \T-atom in \vi into a fresh atomic
proposition is the {\em Boolean abstraction} of \vi, and \vi is the
{\em refinement} of \vip.
Notationally, we indicate by \vip and \mup the Boolean abstraction 
of \vi and $\mu$, and by \vi and $\mu$ the refinements of \vip and
\mup respectively. 
\ignoreinshort{%
With a little abuse of notation, we say that \mup 
is \T-(un)satisfiable iff $\mu$ 
is \T-(un)satisfiable.
}
%
% \ignore{%%%%%%%%%%%%% GIA' DETTO
% %\marg{terminology}
% From now on the following terminology and notation are adopted.
% A superscripted formula $\vip$ is used for denoting the {\it Boolean
%   abstraction}  
% of a \smt formula \vi, which maps Boolean variables 
% into themselves and theory \T-atoms into fresh Boolean atoms
% and distributes with sets and Boolean connectives. 
% (\vi is also said to be the {\em refinement} of \vip.)
% }%%%%%%%%%%%%%%%%%%%
\ignoreinshort{%
We say that  the truth assignment $\mu$ 
\emph{propositionally satisfies} the formula $\varphi$, 
written $\mu \pmodels \varphi$, if $\mu^p \models \varphi^p$.
}

%%\marg{schema of \\lazy SMT}
%\todo{vedi anche descrizione  in papero Silvia\\}
In a lazy \smtt{} solver, %(a complete set of) 
% the truth assignments for  $\vi$ are
% enumerated and checked for \T-satisfiability, returning 
% either  \satres if one \T-satisfiable truth assignment is
% found, \unsatres otherwise.  
%
%In practical implementations, 
the {Boolean abstraction}
\vip of the input formula \vi 
%(obtained by mapping each \T-atom to a fresh atomic proposition)
is given as input to a CDCL SAT solver, and 
whenever a satisfying assignment $\mup$ is found s.t. $\mup\models  \vip$,
the corresponding set of \T-literals $\mu$ is fed to the \Tsolver;
if $\mu$ is found \T{}-consistent, then $\vi$ is
\T{}-consistent; otherwise, 
\Tsolver{} returns a \T-conflict set $\eta$ causing the
inconsistency, so that the clause $\neg\etap$ 
%(the Boolean abstraction of $\neg\eta$) 
is used to drive 
the backjumping and learning mechanism of the SAT solver.
The process proceeds until either a \T{}-consistent assignment $\mu$
is found (\vi is \T-satisfiable), 
or no more assignments are available
(\vi is \T-unsatisfiable).

%%\marg{early-pruning and \\ \T-propagation}
%
Important optimizations are \emph{early pruning} and
\emph{\T-propagation}. The \Tsolver is invoked also when an assignment
$\mu$ is still under construction: if it is \T-unsatisfiable, then the
procedure  backtracks, without exploring the (possibly %up to
many) extensions of $\mu$;
%since no extension of $\mu$ can be \T-satisfiable;
if it is \T-satisfiable, and if the \Tsolver is able to perform  a \T-deduction
$\{l_1,...,l_n\}\models_{\T} l$, 
then $l$ can be unit-propagated, and the \T-deduction 
clause  $(\bigvee_{i=1}^n
\neg l_i \vee l)$ can be  used in backjumping and learning. 
%\RSCHANGEFOUR{
To this extent, in order to maximize the efficiency, most \T-solvers are
{\em incremental} 
and {\em backtrackable}, that is, they are called via a push\&pop interface,
maintaining and reusing the status of the search from one call and the
other. 
%}

%\todo{esempio completo?}
%%% Pure-literal Filtering
%%%%%%%%%%%%%%%%%%%%%%%%%%%%%%%%%%%%%%%%%%%%%%%%%%%%%%%%%%%%%
%\optimization{Pure-literal Filtering}
%The idea behind Pure Literal Filtering is that, 
%%\marg{pure-literal\\ filtering} 
%
\ignoreinshort{
Another optimization is {\em pure-literal filtering}: if some
\larat-atoms occur only positively [resp. negatively] in the original
formula (learned clauses are ignored), then we can safely drop
every negative [resp. positive] occurrence of them from the assignment
$\mu$ to be checked by the \Tsolver \cite{sebastiani07}.
Intuitively, since such occurrences play no role in satisfying the
formula, {the resulting partial assignment ${\mup}'$ still satisfies
  $\vip$.}
The benefits of this action are twofold:
(i) it reduces the workload for the \Tsolver by feeding to it smaller sets;
(ii) it 
increases the chance of finding a \T-consistent satisfying
  assignment by removing   ``useless'' \T-literals
  which may cause the \T-inconsistency of $\mu$.
}

The above schema is a coarse abstraction of the
procedures underlying most
state-of-the-art  \smt{} tools.
The interested reader is pointed  to, e.g., 
\cite{nieot-jacm-06,sebastiani07,BSST09HBSAT}
for details. % and further references.

\subsection{Optimization Modulo Theories}
\label{sec:background_omt}
%\RSTODO{riporta definizione di \omlaratplus}
%
We recall the basic ideas about \omlaratplus and about
 the inline  procedure
in \cite{st-ijcar12,st_tocl14}. 
In what follows, \T is some stably-infinite theory with equality s.t.
\larat and \T are  
 signature-disjoint. 
(\T can be a combination of theories.)
We call an {\em Optimization Modulo $\larat\cup\T$ problem,
  \omlaratplus}, a pair $\pair{\varphi}{\cost}$ such that $\varphi$ is
an \smtlaratplus formula and $\cost$ is an \larat variable occurring in
\vi, representing the cost to be minimized.
The problem consists in finding a \larat-model \calm
for \vi (if any) whose value of \cost  is minimum.
We call an {\em Optimization Modulo $\larat$ problem 
  (\omlarat}) an \omlaratplus problem where \T is empty.
If $\vi$ is in the form $\vi'\wedge\ubliti{c}$ [resp. $\vi'\wedge
\lbliti{c}$] for some value $c\in\mathbb{Q}$, then we call $c$
an {\em upper bound} [resp. {\em lower bound}] for \cost.
If \ub [resp. \lb] is the minimum upper bound [resp. the maximum lower
bound] for \vi,
we also call the interval \range the  {\em range} of \cost.~%

\begin{remark}
\label{remark:multipletheories}
\cite{st-ijcar12,st_tocl14} explain a general technique 
to pass from \omlarat to \omlaratplus by exploiting the 
Delayed Theory Combination technique \cite{bozzanobcjrrs06}
implemented in \mathsatfive.
It is easy to see that this holds also for \laint and \la.  
Therefore, for the sake of brevity and readability, hereafter we consider
the case where \T is the empty theory (\omlarat, \omlaint or \omla),
referring the reader to \cite{st-ijcar12,st_tocl14} for a detailed
explanation about how to handle the general
case.  
\end{remark}

%
%%%%%%%%%%%%%%%%%%%%%%%%%%%%%%%%%%%%%%%%%%%%%%%%%%%%%%%%%%%%%
%%% 
%%%%%%%%%%%%%%%%%%%%%%%%%%%%%%%%%%%%%%%%%%%%%%%%%%%%%%%%%%%%%

%\marg{description of \\inline schema}
In the inline \omlarat schema, 
the procedure takes as input a pair  $\pair{\varphi}{\cost}$, 
plus optionally values for \lb and \ub 
(which are implicitly considered to be $-\infty$ and $+\infty$ if not
present), and returns the model \calm of minimum cost and its 
cost $\currub\defas\calm(\cost)$; it returns the value \ub and an
empty model if \vi is \larat-inconsistent.~%
\ignoreinshort{%
Notice that by providing a lower bound \lb [resp. an upper bound \ub] 
the user implicitly assumes the responsibility of asserting there is no model 
whose cost is lower than \lb [there is a model whose cost is \ub].
}
The standard CDCL-based schema of the SMT solver is modified as follows.

% the whole optimization procedure is pushed
% inside the SMT solver by embedding
% the range-minimization loop inside the CDCL Boolean-search loop of the
% standard lazy SMT schema\ignoreinshort{\ of
%   \sref{sec:background_smt}}. 
%  The SMT solver, which is thus called only once, is modified as follows.
 % \todo{girala come ``modifiche a standard SMT''\\}

\ignoreinshort{
\begin{figure}[th]
  \centering
\scalebox{.29}{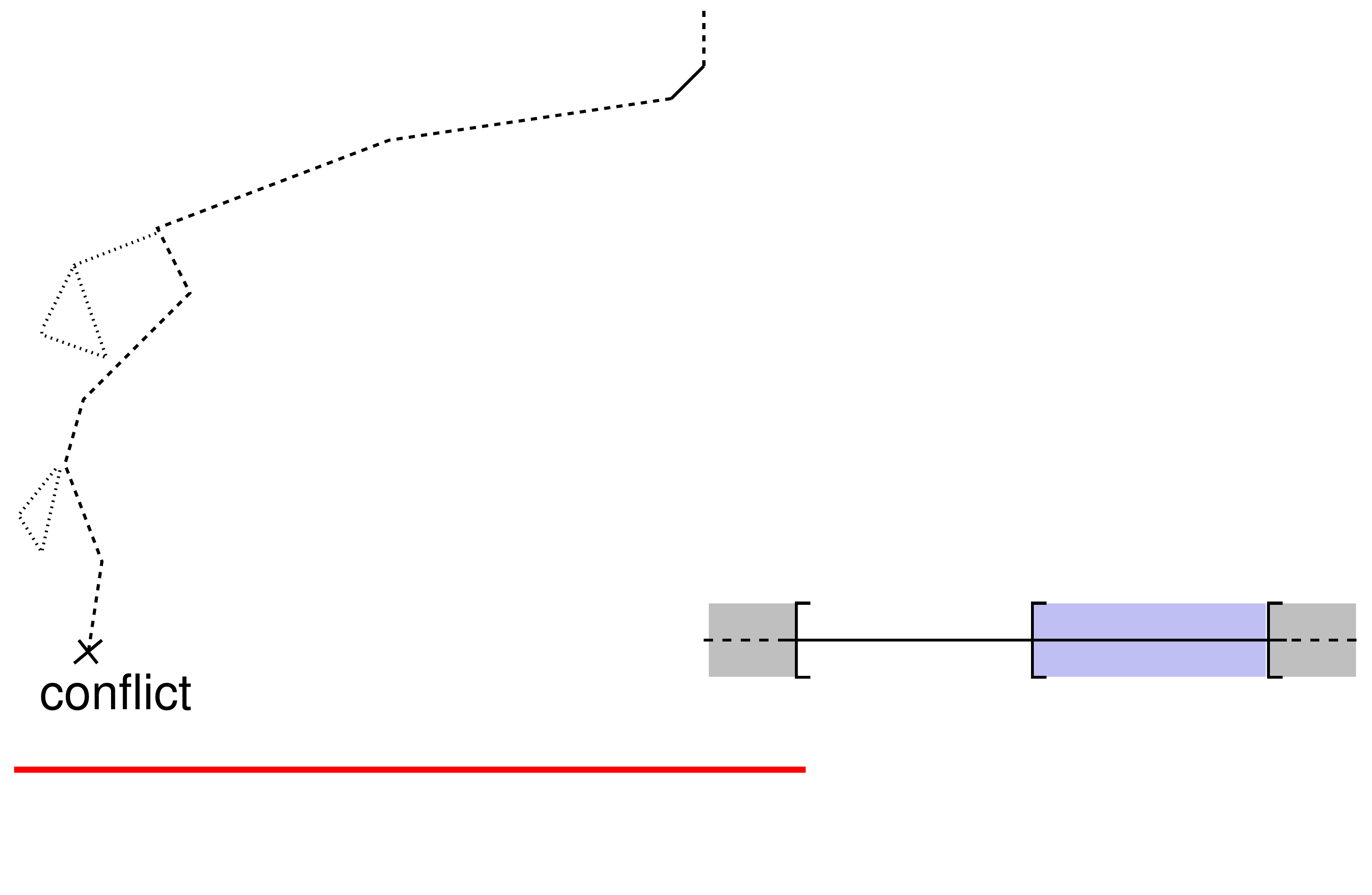} 
\scalebox{.29}{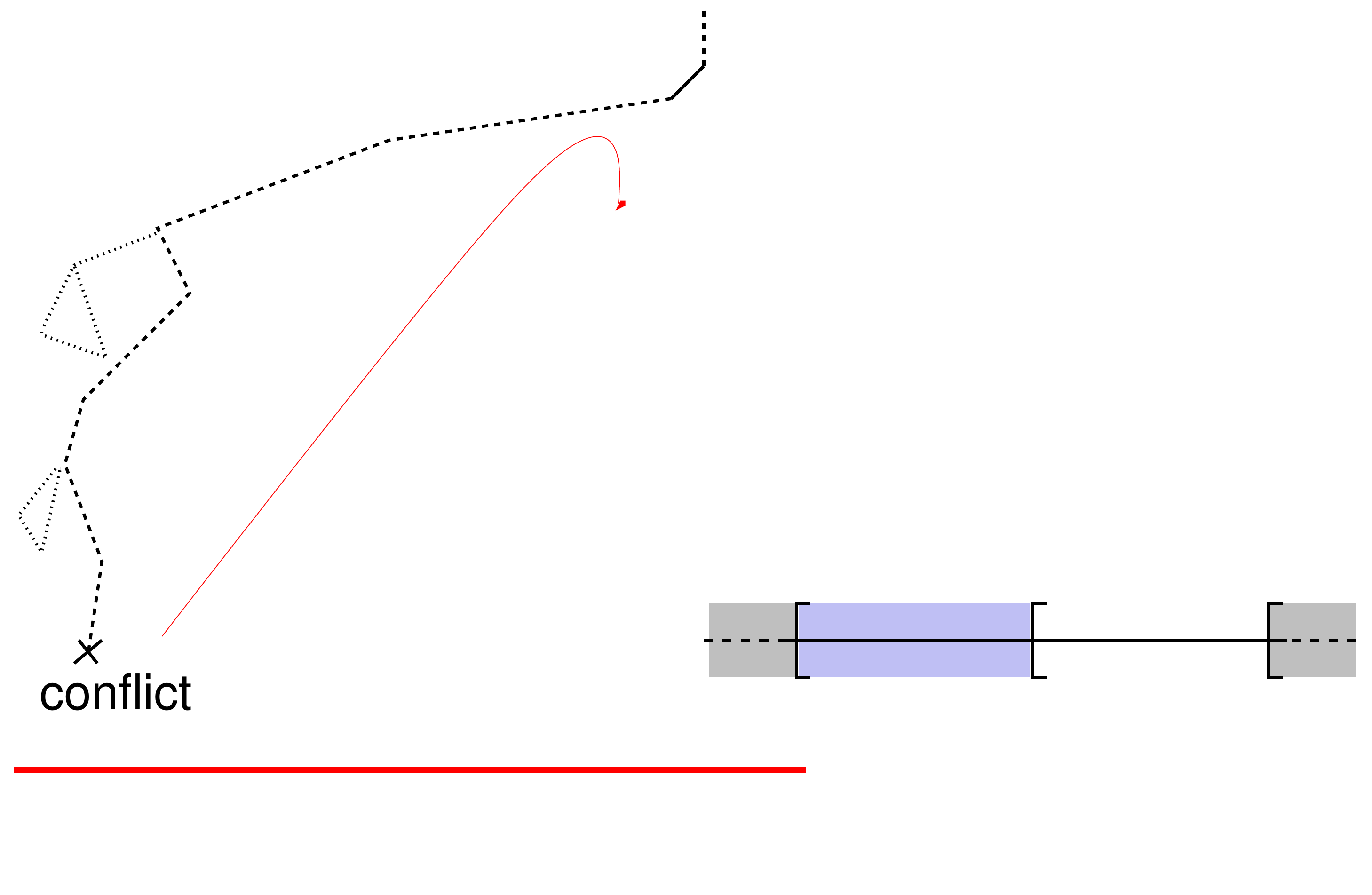} 
%\scalebox{.25}{\input{bin_2.pstex_t}}
%\scalebox{.29}{\input{bin_3.pstex_t}} &
 \scalebox{.29}{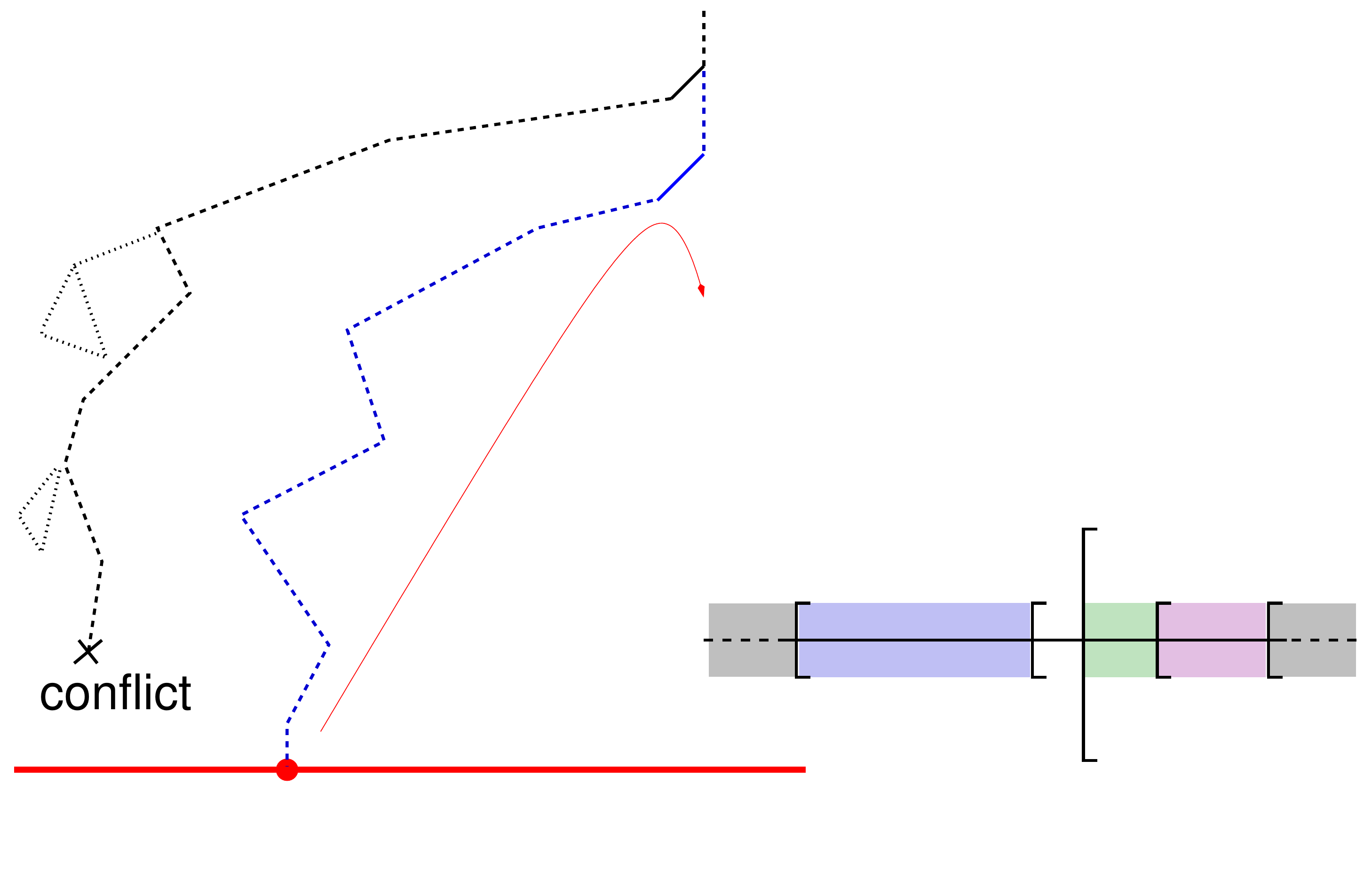}
% \scalebox{.28}{\input{bin_5.pstex_t}}
  \caption{\ignoreinshort{One piece of possible execution of an inline procedure. 
  (i) Pivoting on $(\cost < \pivot_0)$. 
  (ii) Increasing the lower bound to $\pivot_0$.
  (iii) Decreasing the upper bound to $\mincost(\mu_i)$.
  \label{fig:algo_inline}
}}
\end{figure}
}

%
% Moreover, the range-minimization loop  is embedded inside the
% CDCL Boolean-search loop of the standard lazy SMT schema in
% \sref{sec:background_smt}.  
%  (In a
% nutshell, as soon as a new minimum-cost model of cost \currub is found, the unit
% clause \currublit is learned, which causes the CDCL SAT solver 
% to backtrack to level 0 and to restart the search from there.)
%
% A much more sophisticated version is implemented by embedding the
% minimization loop  inside the Boolean search performed by the 
% CDCL-based \smt solver. 
% (We call {\em inline} this variant and {\em offline} that of
% Algorithm~\ref{algo:mixsearch}.)
%This work as follows.

\noindent {\bf Initialization.}
the variables \currlb, \currub (defining the current range) 
 are initialized to \lb and \ub respectively, the variable \pivot
(defining the pivot in binary search)
 is not initialized,  
the \larat-atom \pivotatom is initialized to $\top$
and the output model \calm is initialized to be an empty model.

\noindent{\bf Range Updating \& Pivoting.}
Every time the search of the CDCL SAT solver gets back to decision level 0,
the range \currrange is updated s.t. 
\currub [resp. {\currlb] is assigned the lowest [resp. highest] value
  $\currub_i$ [resp. $\currlb_i$] such that the atom $(\cost <
  \currub_i)$ [resp. $\neg (\cost < \currlb_i)$] is currently assigned
  at level 0.
%
%\noindent{\bf Pivoting.}
Then the heuristic function \dopivoting{} is invoked, which 
decides 
whether to run the current step in binary- or in linear-search mode:
in the first case (which can occur only if $\currlb>-\infty$ and $\currub<\infty$) a value
$\pivot\in\ 
]\currlb,\currub[$ is computed (e.g. $\pivot=(\currlb+\currub)/2$), and the (possibly new) atom $\pivotatom\defas
(\cost < \pivot)$ is decided to be true (level 1) by the SAT solver.  
This
% mimics steps \ref{algomix:initpivoting}-\ref{algomix:addpivot}
% in Algorithm~\ref{algo:mixsearch},
 temporarily restricts the cost range to
  $[\currlb,\pivot[$.  
Then the CDCL solver proceeds its search, as in \sref{sec:background_smt}.

% In either case, then the standard
% lazy SMT search proceeds, until one of the following facts happens:
%

\noindent{\bf Decreasing the Upper Bound.} 
When an assignment $\mu$  is generated s.t. $\mup\models\vip$ and 
which is found \larat-consistent by \laratsolver, $\mu$ is also fed to
\laratminimize, returning the minimum cost \minvalue of $\mu$; then the unit
clause $C_{\mu}\defas(\cost < \minvalue)$ is learned and fed to the
backjumping mechanism, which forces the SAT solver to backjump to
level 0, then unit-propagating $(\cost < \minvalue)$. This  restricts the cost range
to $[\currlb,\minvalue[$.
\laratminimize is embedded within \laratsolver{} --it is a simple extension of
the LP algorithm in \cite{demoura_cav06}--  so that it is
 called incrementally after it, without restarting its search from scratch.  
%\todo{espandi}
Notice that the clauses $C_{\mu}$ ensure progress in the minimization 
every time that a new \larat-consistent assignment is generated. 

\noindent{\bf Termination.} %
The procedure terminates when 
%either $\currub \le \currlb$ or 
the
embedded SMT-solving algorithm reveals an inconsistency, 
returning the current values of \currub and \calm.

As a result of these modifications, we also have the following typical
scenario\ignoreinshort{\ (see Figure~\ref{fig:algo_inline})}.

%\marg{right part}
\noindent{\bf Increasing the Lower Bound.} 
In binary-search mode, when a conflict
occurs and the conflict analysis of the SAT solver produces a conflict clause in the
form $\neg\pivotatom\vee\neg\eta'$ s.t. all literals in $\eta'$ are
assigned \trueval{} at level 0 (i.e., $\vi\wedge\pivotatom$ is
\larat-inconsistent), then the SAT solver backtracks to level 0,
unit-propagating $\neg\pivotatom$.  This case  permanently restricts the cost range
to $[\pivot,\currub[$.

%\ignoreinshort{Notice that this procedure maintains the invariant that
%$(\cost<\currub)\in \mu$.}
Notice that, to guarantee termination, 
{binary-search steps must be interleaved with linear-search ones infinitely often.}
%occur infinitely often.
We refer the reader to \cite{st-ijcar12,st_tocl14} for details and 
for a description of further improvements to the basic inline procedure.
% \ignoreinlong{
% Some improvements to the basic procedure are described 
% in \cite{st-ijcar12,st_tocl14}.}
\ignoreinshort{
% First, an early-pruning call to \laratsolver is performed also at level 0,
% before the search starts and/or \pivotatom is decided, so that 
\ignore{%
First, an early-pruning call to \laratsolver is performed
  also at level 0, before the first decision is performed, so that to
  cut the search if the current set of unit clauses is
  \larat-inconsistent.%
}
}

%%%%%%%%%%%%%% END IGNORE

\section{Pushing the envelope of OMT}
\label{sec:pushing}

\subsection{From \omlarat to \omla}
\label{sec:omtlaint_proc}
%\RSTODO{PARLARE da \laint a \la\\}
%\RSTODO{togliere picot\\}
% We start from the observation that the only \larat-specific
% components of the online \omlarat schema of \sref{sec:background_omt}
% is the fact that \larat-specific \T-solving and minimizing procedures
% are required.
{%
We start from the observation that the only \larat-specific
components of the inline \omlarat schema of \sref{sec:background_omt}
are the \T-solving and minimizing procedures.}
Thus, under the assumption of having an efficient \lasolver{}  already 
implemented inside the embedded SMT solver --like we  have 
in \mathsatfive \cite{griggio-jsat11}--
the schema in \sref{sec:background_omt} can be adapted to \la by
 invoking an \la-specific
minimizing procedure each time a truth-assignment $\mu$
s.t. $\mu^p\models\varphi^p$ is generated.

\begin{remark}
Notice that in principle in \laint the minimization step is not
strictly necessary if the input problem is lower bounded.  In fact, to
find the optimum $\cost$ value it would be sufficient to iteratively
enumerate and remove each solution found by the standard
implementation of the \laintsolver{}, because each step guarantees an
improvement of at least 1.
Minimizing the $\cost$ value at each iteration of the \smt engine,
however, allows for speeding up the optimization search by preventing
the current truth assignment $\mu$ from being generated more than once. In
addition, the availability of a specialized \laintminimize procedure
is essential to recognize unbounded problems.
\end{remark}

The problem of implementing an efficient \omla tool
 reduces thus to that of implementing an efficient 
minimizer in \la, namely \laminimize,
{\em which exploits and cooperates in synergy 
 with the other components of the \smt solver}.
In particular, it is advisable that \laminimize is embedded into the \lasolver,
 so that it is called
incrementally after the latter has checked the \la-consistency of 
the current assignment $\mu$. 
%and returned the current (non minimum) model \calm. 
(Notice that, e.g., embedding into \laminimize 
a MILP tool from the  shelf would not match these requirements.)
%
%\ignore{%%%%%%%% TOLGO?
%\RSTODO{Questa notazione e` necessaria? check e se no semplifica}
%In our problem domain,
%$\cost=\underset{x\in\mathcal{X}}{\mbox{min}}\:c^\top x$.  
%}
%
% NOTA: citazione RUIP ``Recognizing Unbounded Integer Programs'' by
% R.H.Byrd, A.J.Goldman, Miriam Heller - rimossa in quanto non
% rilevante
%
%\smallskip
To this extent, 
% In order to adopt an 
% optimization scheme that integrates well with the pre-existing
% implementation of the \lasolver and the rest of the lazy SMT-solving
% schema in \optimathsat, 
we have  investigated  both theoretically and empirically three
different schemas of Branch\&Bound \laminimize procedure, 
which we call {\em basic}, {\em advanced} and {\em truncated}.

The first step performed by \laminimize  is to check
whether $\cost$ is lower bounded.  
Since a {\em feasible} \textit{MILP} problem is unbounded if and
only if its corresponding continuous relaxation  is
unbounded \cite{byrd_unboundedMILP},%
\footnote{As in \cite{byrd_unboundedMILP}, by ``continuous
  relaxation'' --henceforth simply ``relaxation''--
we mean that the integrality constraints on the
  integer variables are relaxed, so that they can take fractional
  values.}
we run \laratminimize  on the
relaxation of $\mu$.
%
% $\underset{x\in\mathcal{P}'}{\mbox{min}}\:c^\top x$ to get
% a \textit{fractional solution} ${x}^*_{LP}\in\mathbb{R}^n$.
If the relaxed problem if unbounded, then \laintminimize{} 
returns $-\infty$; 
otherwise, \laratminimize returns the minimum value of \cost in the
relaxed problem, which we set as the current {\em lower bound} \lb for
\cost in the original problem. 
We also initialize the {\em upper bound} \ub for \cost to the value
$\mathcal{M}(\cost)$, where $\mathcal{M}$ is the 
model returned by the most recent call to the \lasolver{} on $\mu$.

Then we explore
the solution space  by means of an LP-based Branch\&Bound
procedure that reduces the original MILP problem to a sequence of
smaller sub-problems, which are solved separately.

\fakesubsubsection{Basic Branch\&Bound}
%
% \RSTODO{you might want to erase 3rd paragraph and second sentence in 1st paragraph to spare space
% since we are over 2 pages.. these parts are not fundamental.}
%
We describe first 
a naive version of the Branch\&Bound minimization procedure.
(Since it is  very inefficient, we present it only as a
 baseline for the other approaches.)
% GIA? DETTO PRIMA
% We start by initializing the optimization search upper bound
% $ub$ to $\mathcal{M}(\cost)$ value, where $\mathcal{M}$ is the
% \la{}-model that satisfies $\varphi$ which has been found in the
% most recent call to the \lasolver{}.
%
We first invoke \laratminimize{} on the relaxation of the 
current \la problem.
If the relaxation is found \larat-unsatisfiable, then 
also the original problem is \la-unsatisfiable, and the procedure
backtracks. 
Otherwise, \laratminimize returns a minimum-cost model \calm 
of cost \minvalue. If such solution is \la-compliant, 
then we can return \calm and \minvalue,
setting $\ub=\minvalue$.~%
(By ``\la-compliant solution'' here we mean 
that the integer variables are all given integer values, whilst
rational variables can be given fractional values.)
 % \PTTODO{Patrick verifica la
 %  footnote. Verifica anche che che la parola
 %  ``\la-compliant'' in quanto segue sia sempre usata a proposito.} 

Otherwise, we select an integer variable $x_j$ which is given a
fractional value $x^*_j$ in \calm as \textit{branching variable}, 
and split the current problem into a
pair of complementary sub-problems, 
by augmenting them respectively with the linear cuts 
$(x_j \leq \lfloor x^*_j\rfloor)$ and
$(x_j \geq \lceil x^*_j\rceil)$.  Then, we separately explore
each of these two sub-problems in a recursive fashion, and we return
the best of the two minimum values of $\cost$ which is found in the two
branches, with the relative model.

In order to make this exploration more efficient, as the recursive
Branch\&Bound search proceeds, we keep updating the upper bound  $\ub$
to the current best value of $\cost$ corresponding to an \la-compliant
solution. 
Then, we can prune all sub-problems in which the \larat{}
optimum $\cost$ value is greater or equal than $\ub$, as they cannot
contain any better solution.

\fakesubsubsection{Advanced Branch\&Bound}
Unlike the basic scheme, the advanced Branch\&Bound is built on top
of the \lasolver{} of \mathsatfive and takes advantage of all the
advanced features for performance optimization that are already
implemented there \cite{griggio-jsat11}. 
In particular, we re-use  its very-efficient 
 {internal Branch\&Bound} procedure for \la-solving, which 
exploits historical information to drive the 
search and achieves higher pruning by {\em back-jumping} within the
Branch\&Bound search tree, driven by the analysis of
unsatisfiable cores. (We refer the reader to \cite{griggio-jsat11} for details.)

We adapt the \la-solving algorithm of \cite{griggio-jsat11} to
minimization as follows.
As before, the minimization algorithm
 starts by setting $\ub=\mathcal{M}(\cost)$, %where
$\mathcal{M}$ being the model for $\mu$ which was returned by the most
recent call to the \lasolver{}.  
Then the linear cut $(\cost < \ub)$ is pushed on top of the
constraint stack of the \lasolver{}, which forces the search to look for a
better \la-compliant solution than the current one. 

Then, we
use the {internal Branch\&Bound} component of the
\lasolver{} to seek for a new \la-compliant solution.  
The first key modification  is that we invoke \laratminimize on each node of
Branch\&Bound search tree to ensure that $x_{LP}^*$ is optimal in the
\larat domain.
The second modification is that,  
 every time a new  solution  is found --whose cost  $\ub$ improves the
 previous upper bound by construction-- we 
empty the stack of 
\lasolver, push there a new cut in the form $(\cost < \ub)$ and
restart the search.
 Since the problem is known to be bounded, there are 
only a finite number of \la-compliant solutions possible that can be removed from
the search space. Therefore, the set of constraints is guaranteed to
eventually become unsatisfiable, and at that point $\ub$ is returned
as optimum $\cost$ value in $\mu$ to the SMT solver, which learns 
the unit clause $C_{\mu}\defas(\cost < \ub)$.

\fakesubsubsection{Truncated Branch\&Bound}
%
% NOTA x RS:
% - in realtà - come dicevi tu - l'overhead c'è (sufficiente che il taglio della soluzione ottima 
%   restituisca Politopo non vuoto), ma in generale è contenuto.
% - il vero vantaggio di tale strategia sta nell'usare metodi ILP avanzati per problemi degeneri
%   per i quali il metodo del simplesso non è efficace. Infatti, possiamo decidere di interrompere 
%   la internal Branch\&Bound di Griggio anche quando superiamo un certo numero di passi.
%   Questa cosa non è menzionata per questioni di semplicità espositiva nella precedente esposizione
%
%
We have empirically observed that in most cases the above scheme is
effective enough that a single loop of
{advanced Branch\&Bound} is sufficient to find
the optimal solution for the current truth assignment
$\mu$. However, the advanced Branch\&Bound procedure still performs an
additional loop iteration to prove that such solution is indeed
optimal, which  causes
additional unnecessary overhead.  Another drawback of {advanced} B\&B
 is that for degenerate problems the
Branch\&Bound technique is very inefficient. In such cases, it is more
convenient to interrupt the B\&B search and simply return
$\ub$ to the \smt{} solver, s.t. the unit clause $C_{\mu}\defas(\cost < \ub)$
is learned; in fact,  in this way we can easily re-use
the entire stack of \lasolver{} routines in \mathsatfive{} to find an
improved solution more efficiently.
 
Therefore, we have implemented a ``sub-optimum'' 
variant of \laminimize in which the
inner \lasolver{} minimization procedure stops as soon as either it finds
its first solution or it reaches a certain limit on the number of
branching steps. The drawback of this variant is that, in some cases,
it analyzes a truth assignment $\mu$ (augmented with the extra constraint
 $(\cost < \ub)$) more than once.

\subsection{Multiple-objective OMT}
\label{sec:multiomt_proc}
%TODO{Sistemare la questione del dominio reali vs interi.\\}
% TODO{Dire qualcosa sulla possibilita' di aggiungere lower e
%   upper bound?\\}
%
%\noindent
We generalize the \omla problem to multiple cost functions as follows.
\ignoreinshort{(As with plain \omla, the extension to \omlaplus
  follows the 
technique described in \cite{st-ijcar12,st_tocl14}.)}
A {\em multiple-cost \omla problem} is a pair $\pair{\varphi}{\calc}$
s.t $\calc \defas \{\cost_1,...,\cost_k\}$ is a set of \la-variables occurring
in \vi, and consists in finding a set of \la-models 
$\{\calm_1,...,\calm_k\}$ s.t. each $\calm_i$ makes
$\cost_i$ minimum.
\localbounds{%%%%%%%%%%%%%%%%%%% BEGIN LOCAL & GLOBAL BOUNDS
If $\vi$ is in the form:
\begin{equation}
\label{eq:multiplebounds}
\vi'\wedge
\bigvee_{\cost_i\in\C} ((\cost_i<\ubgen{}{i}) \wedge \neg(\cost_i<\lbgen{}{i}))
\end{equation}
\globalbounds{
\begin{equation}
\label{eq:multiplebounds}
\vi'\wedge
\bigvee_{\cost_i\in\C} (\cost_i<\ubgen{}{i}) 
\wedge  
\bigwedge_{\cost_i\in\C} \neg(\cost_i<\lbgen{}{i}), 
\end{equation}
}
then $\ubgen{}{i}\in\mathbb{Q}\cup\{+\infty\}$ 
[resp. $\lbgen{}{i}\in\mathbb{Q}\cup\{-\infty\}$]
is called the {\em upper bound} [resp. the {\em lower bound}] of
$\cost_i$ in \vi.
\footnote{This definition is compliant with that of \cite{li_popl14} --where 
objective functions $k_i$ are {\em maximized} and it is not provided
any explicit notion of bounds:
we can translate a problem from \cite{li_popl14} by setting
$\cost_i \defas -k_i$ $\ubgen{}{i}\defas+\infty$ and $\lbgen{}{i}\defas-\infty$ for
each i; the viceversa can be obtained by inserting the bound constraints
inside the formula as in \eqref{eq:multiplebounds}.}
(Here we implicitly assume that every atom in the form 
$(\cost_i<+\infty)$
  [resp. $(\cost_i<-\infty)$] is simplified into $\top$  [resp. $\bot$]
  and that \eqref{eq:multiplebounds} is simplified accordingly.)~%
}
\globalbounds{
Typically, an upper bound \ubgen{}{i} is added when we know there exist a
\la-model \calm of \vi s.t. $\calm(cost_i)=\ubgen{}{i}$;
vice versa,  a lower bound \lbgen{}{i} is added when we know there exist no
\la-model \calm of \vi s.t. $\calm(cost_i)<\lbgen{}{i}$.~%
\footnote{
Notice that if $\ubgen{}{i}=+\infty$ for some $i$, then the whole
disjunction in \eqref{eq:multiplebounds} simplifies into $\top$.
Typically the upper bounds come from the fact that we know (at least) 
one model \calm for \vi, so that $\ubgen{}{i}\defas\calm(\cost_i)$ for
every $i$.}
}%%%%%%%%%%%%%%%%%%% END LOCAL & GLOBAL BOUNDS
%
%
%%%%%%%%%%%%%%%%%%%%%%%%%%%%%%%%%%%%%%%%%%%%%%%%%%%%%%%%%%%%%
%%% Procedure
%%%%%%%%%%%%%%%%%%%%%%%%%%%%%%%%%%%%%%%%%%%%%%%%%%%%%%%%%%%%%
%\marg{Procedura}
We extend the \omlarat [\omla] procedures of
\sref{sec:background_omt} and \sref{sec:omtlaint_proc} 
to handle multiple-cost problems. The procedure works in linear-search
 mode only.%
\ignoreinlong{
\footnote{
  Since the linear-search versions of the
  procedures in \sref{sec:background_omt} and \sref{sec:omtlaint_proc} 
differ only for the fact that they invoke \laratminimize and
\laminimize respectively, here we do not distinguish between them.
We only implicitly make the assumption that the
\laminimize does not work in truncated mode, so that it is
guaranteed to find a minimum in one run. Such assumption is not
strictly necessary, but it makes the explanation easier.
}}
\ignoreinshort{\begin{remark}
  Since the linear-search versions of the
  procedures in \sref{sec:background_omt} and \sref{sec:omtlaint_proc} 
differ only for the fact that they invoke \laratminimize and
\laminimize respectively, here we do not distinguish between them.
We only implicitly make the assumption that the
\laminimize does not work in truncated mode, so that it is
guaranteed to find a minimum in one run. Such assumption is not
strictly necessary, but it makes the explanation easier.
\end{remark}}
It  takes as input a pair
$\pair{\varphi}{\calc}$  and returns
a list of minimum-cost models $\{\calm_1,...,\calm_k\}$, 
plus the corresponding list of minimum values 
$\{\currub_1,...,\currub_k\}$. (If \vi is \la-inconsistent, it
returns $\currub_i=+\infty$ for every $i$.)

\noindent {\bf Initialization.}
First,  we set $\currub_i=+\infty$  for every $i$, and we
set $\currentcosts=\calc$, s.t. \currentcosts{} is the list of
currently-active cost functions. 

\noindent{\bf Decreasing the Upper Bound.} 
When an assignment $\mu$  is generated s.t. $\mup\models\vip$ and 
which is found \la-consistent by \lasolver, $\mu$ is also fed to
\laminimize. 
For each $\cost_i\in\currentcosts$:
\begin{renumerate}
  \item  \laminimize finds an \la-model $\calm$ for
$\mu$ of minimum cost $\minvalue_i$;

  \item if $\minvalue_i$ is $-\infty$, then there is no more reason to
    investigate $\cost_i$, so that we set $\currub_i=-\infty$
    and $\calm_i=\calm$, and $\cost_i$ is dropped from
    $\currentcosts$;
  \item if $\minvalue_i<\currub_i$, then we set
    $\currub_i=\minvalue_i$ and $\calm_i=\calm$.
\end{renumerate}
As with the single-cost versions\ignoreinshort{\ of \sref{sec:background_omt}}, 
\laminimize is embedded within \lasolver,  so that it is
 called incrementally after it, without restarting its search from scratch.
After that, the  clause 
\begin{equation}
\label{eq:multipleblockingclause}
C_{\mu}\defas  \bigvee_{\cost_i\in\currentcosts} (\cost_i<\currub_i)
\end{equation}
is learned, and the CDCL-based SMT solving process proceeds its
search. Notice that, since by construction $\mu\wedge
C_{\mu}\models_{\la}\bot$, a theory-driven backjumping step \cite{BSST09HBSAT}
will occur as soon as $\mu$ is extended to assign to true some literal
of   $C_{\mu}$. 

\noindent{\bf Termination.} The procedure terminates either when 
$\currentcosts$ is empty or when 
\vi is found \la-inconsistent.
(The former case is a subcase of the latter, because 
it would cause the generation of an empty clause $C_{\mu}$
\eqref{eq:multipleblockingclause}.) 

%\smallskip
The clauses $C_{\mu}$ \eqref{eq:multipleblockingclause} ensure a progress in the
minimization of one or more of the $\cost_i$'s 
every time that a new \la-consistent assignment is generated.
We notice that,  by construction, $C_{\mu}$
 is such that $\mu\wedge
C_{\mu}\models_{\la}\bot$,  so that each $\mu$ satisfying the
original version of $\vi$ 
can be investigated by the minimizer only once. 
 Since we have only a finite number of such candidate
assignments for $\vi$, this guarantees the termination of the
procedure. The correctness and completeness is guaranteed by these of
\laminimize{}, which returns the minimum values for each such assignment.

%\smallskip
%%%%%%%%%%%%%%%%%%%%%%%%%%%%%%%%%%%%%%%%%%%%%%%%%%%%%%%%%%%%%
%%% Example 
%%%%%%%%%%%%%%%%%%%%%%%%%%%%%%%%%%%%%%%%%%%%%%%%%%%%%%%%%%%%%
%%\marg{Esempio} 
To illustrate the behaviour of our procedure,
and to allow for a direct comparison wrt. the procedure described in
\cite{li_popl14}, in Figure~\ref{ex:multipleomt} we present its
execution  on the toy example \larat-problem in \cite{li_popl14}.
Notice that, unlike the algorithm in \cite{li_popl14},
 our procedure is driven by the Boolean search:
 each time a novel assignment is generated, it eagerly 
produces the maximum progress for as many $\cost_i$'s as
possible. 
%
%\ignoreinshort{\PTTODO{Eventualmente amplia.}
The algorithm described in \cite{li_popl14}, instead,
 does not use a LP minimization
 procedure: rather, a sequence of blackbox calls to an underlying SMT
solver (\zthree) allows for finding progressively-better solutions along some
objective direction, either forcing discrete jumps to some bounds induced by the
inequalities in the problem, or proving such objective is unbounded. 
%}

\newcommand{\exformula}{\ensuremath{
    \begin{array}{lll}
\vi&\defas&(1\le y)\wedge
(y\le 3)\\
&\wedge&
(((1\le x)\wedge(x\le 3))\vee(x\ge 4))\\
&\wedge& (\cost_1=-y)\wedge(\cost_2=-x-y)
      
    \end{array}
}}
\newcommand{\exmuone}{\ensuremath{
    \begin{array}{lll}
\mu_1&\defas& \{(1\le y),(y\le 3),(1\le x),(x\le 3),\\
&&(\cost_1=-y),(\cost_2=-x-y)\}      
    \end{array}
}}
\newcommand{\exmutwo}{
\ensuremath{
    \begin{array}{lll}
\mu_2&\defas& \{(\cost_1=-y),(\cost_2=-x-y),\\
& & (1\le y),(y\le 3),(x\ge 4),\\
& &    (\cost_2<-6)\}
    \end{array}
}}
\begin{figure}[t]
\scalebox{1.2}{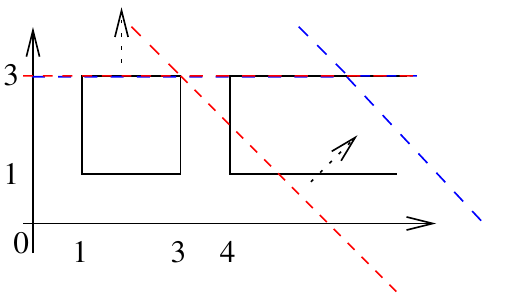}
  \caption{\label{ex:multipleomt}
  In one possible execution over the \larat-formula \vi, the
  CDCL-based SMT engine finds the truth assignment $\mu_1$
  first,   which is
  found \larat-consistent by the \larat-solver.  (For the sake of
  readability, we've removed from the
  $\mu_i$'s the redundant literals like ``$\neg(x\ge 4)$'' from
  $\mu_1$.)  Then the minimizer finds the minima $\minvalue_1=-3$,
  $\minvalue_2=-6$, the upper bounds are updated to these values, and the
  clause $(\cost_1<-3)\vee(\cost_2<-6)$ is learned.  The next
  \larat-consistent assignment found is necessarily $\mu_2$,
from which the minimizer finds the minima $\minvalue_1=-3$,
  $\minvalue_2=-\infty$. Hence $\cost_2$ is dropped from \currentcosts, 
 and the unit clause $(\cost_1<-3)$ is learned, making $\vi$ 
  \larat-inconsistent, so that no more assignment is found and
 the procedure terminates. 
\newline
 In a luckier execution $\mu_2\setminus\{(\cost_2<-6)\}$ is found first, thus the
 minimizer finds directly the minima $\minvalue_1=-3$,
  $\minvalue_2=-\infty$ s.t.  $(\cost_1<-3)$ is learned, and the procedure
  terminates without generating $\mu_1$. 
}
\end{figure}

% TODO{Alcuni cenni di confronto con \cite{li_popl14}?: 
% SAT-driven, e a ogni nuovo truth assignment 
% possible progress in many directions contemporarily.\\}
%\PTTODO{PATRICK, verifica quanto segue:\\}
\smallskip
The procedure {is} improved in various ways. 
%\ignoreinshort{%
%
%}
First, we notice that the clause $C_{\mu}$ is 
strictly stronger than the  clause $C_{\mu'}$ which was generated with the
previous truth assignment $\mu'$, so that $C_{\mu'}$ can be safely
dropped, keeping only one of such clauses at a time.
This is as if we had only one such clause 
whose literals are progressively strengthened.
\ignoreinshort{%
}
Second, before step (i), the constraint $(\cost_i<\currub_i)$ can be temporarily
pushed into $\mu$: if \laminimize returns \unsatres, then 
there is no chance to improve the current value of $\currub_i$, so
that the above
constraint can be popped from $\mu$ and step (ii) and (iii) can be skipped for
the current $\cost_i$.
%\ignoreinshort{%
%
%}
Third, in case the condition in step (iii) holds, it is possible to learn also the \la-valid clause 
$(\cost_i<\currub_i)\imp(\cost_i<\currub_i')$ s.t. $\currub_i'$ is the
previous value of $\currub_i$. This allows for ``activating'' all
previously-learned clauses in the form 
$\neg(\cost_i<\currub_i')\vee C$ 
as soon as $(\cost_i<\currub_i)$ is assigned to true.

% \ignoreinshort{%
% TODO{un paragrafo su upper andllower bounds e su come 
% possano migliorare le performences (ad es se $lb_i>ub_i$ posso
% togliere $\cost_i$ da \currentcosts?}
% }
 
{%
\subsubsection{Lexicographic combination.}
%\paragraph{Lexicographic combination.}
%
As in \cite{bjorner_scss14}, we easily extend our inline procedure
to deal with the {lexicographic combination} of multiple costs 
$\set{\cost_1,...,\cost_k}$. 
\ignoreinshort{This works as follows.}
\ignore{
As soon as we find a minimum $\currub_{1}$ for $\cost_1$ with 
model $\calm_1$
}
We start by looking for a minimum for $\cost_1$:
as soon as a minimum $\currub_{1}$ with its model $\calm_1$ is found, 
if $\currub_{1}=-\infty$ then we stop, otherwise
we substitute inside $\vi$ 
the unit clause $(\cost_1<\currub_1)$ with $(\cost_1=\currub_1)$, 
we set $\currub_2\defas\calm_1(\cost_2)$, 
and we look for the minimum of $\cost_2$ in the resulting formula.
%\ignoreinshort{and so on, until all $\cost_i$'s have been considered.}
This is repeated until all $\cost_i$'s have been considered.
}

\subsection{Incremental OMT}
\label{sec:incomt_proc}
Many modern  SMT solvers, including \mathsatfive, 
provide a {\em stack-based 
  incremental interface}\ignoreinshort{\ (see e.g. \cite{eensorensson:sat2003})}, by
which it is possible to push/pop 
%(the blocks of clauses corresponding to) 
sub-formulas $\phi_i$ into a stack of formulas
$\Phi\defas\set{\phi_1,...,\phi_k}$, and then to check incrementally the
satisfiability of $\bigwedge_{i=1}^k \phi_i$. 
The interface maintains the {\em status} of the search from one call
to the other, in particular it records the {\em learned clauses} (plus other information).
Consequently, when invoked on
$\Phi$, %$\Phi\defas\set{\phi_1,...,\phi_k}$
the solver can reuse a clause $C$ which was learned during a previous
call on some $\Phi'$ %$\Phi'\defas\set{\phi'_1,...,\phi'_{k'}}$
if $C$ was derived only from clauses %in $\Phi'$
which are still in $\Phi$.~%
\ignoreinshort{%%%%%%%%%%%%%%%%%% LONG VERSION
\footnote{%
  Provided $C$ was not discharged in the meantime.%
} 
In particular, if $\Phi'\subseteq\Phi$, then the solver can reuse
all clauses learned while solving $\Phi'$.  
} %%%%%%%%%%%%%%%%%% END LONG VERSION

%TODO{TAGLIARE PROSSIMO PAR?} 

%\cutifnecessary{%
In particular, in \mathsatfive incrementality is achieved by 
first rewriting $\Phi$ into $\set{A_1\imp\phi_1,...,A_k\imp\phi_k}$, each
$A_i$ being a fresh Boolean variable, 
and then by running the SMT solver under the assumption of the 
variables $\{A_1,...,A_k\}$, in such a way that every learned clause 
which is derived from some $\phi_i$ is in the form $\neg A_i\vee C$ 
\cite{eensorensson:sat2003}. Thus it is possible to safely keep the
learned clause from one call to the other because, if $\phi_i$ is
popped from $\Phi$,
then $A_i$ is no more assumed, so that the clause $\neg A_i\vee C$  is
inactive. (Such clauses can be garbage-collected from time to time to
reduce the overhead.) 
%}

%TODO{avere \omt incrementale a' utile perche'...}
Since none of the OMT tools in
\cite{st-ijcar12,st_tocl14,li_popl14,LarrazORR14} provides an
incremental interface, nor such paper explains how to achieve it, 
here we address explicitly the problem of making OMT incremental.

We start noticing that
if (i) the OMT tool is based on the schema %of \cite{st-ijcar12,st_tocl14} 
in \sref{sec:background_smt} or on its \la and multiple-cost
extensions of \sref{sec:omtlaint_proc} and
\sref{sec:multiomt_proc}, and
(ii)  the embedded SMT solver has an incremental interface, like that
of \mathsatfive, 
then an OMT tool can be easily made incremental by exploiting
the incremental interface of its  SMT solver. 

In fact, in our OMT schema all learned clauses are
either \T-lemmas %(and hence they are valid in \larat) 
or they are
derived from \T-lemmas and some of the subformulas $\phi_i$'s, {\em with
the exception of the clauses $C_\mu\defas(\cost<\minvalue)$
(\sref{sec:background_omt}) 
[resp. $C_\mu\defas(\cost<\minvalue)$ (\sref{sec:omtlaint_proc}) and
$C_\mu\defas\bigvee_{\cost_i\in\currentcosts}(\cost_i<\currub_i)$ (\sref{sec:multiomt_proc}),]
} which are ``artificially'' introduced to
ensure progress in the minimization steps.
%\footnote{
(This
  holds also for the unit clauses $(\pivotatom)$ 
which are learned in an improved version, see \cite{st-ijcar12,st_tocl14}.)
Thus, in order to handle incrementality, it suffices to drop only these
clauses from one \omt call to the other,
 while preserving all the others, as with incremental SMT.

%\cutifnecessary{% 
In a more elegant variant of this technique, which
we have used in our implementation, at each incremental call to OMT
(namely the $k$-th call) a fresh Boolean variable $A^{(k)}$ is
assumed.  Whenever a new minimum \minvalue is found, the augmented
clause $C_{\mu}^*\defas\neg A^{(k)}\vee (\cost < \minvalue)$ is
learned instead of $C_{\mu}\defas(\cost < \minvalue)$.  In the
subsequent calls to OMT, $A^{(k)}$ is no more assumed, so that the
augmented clauses $C_{\mu}^*$'s which have been learned during the
k-th call are no more active.  
%}

% \begin{remark}
Notice that in this process 
reusing the clauses that are learned by the underlying SMT-solving
steps is not the only benefit.
In fact also the learned
clauses in the form $\neg(\cost<\minvalue)\vee C$ which may be 
produced after learning $C_\mu\defas(\cost<\minvalue)$ are preserved
to the next OMT calls. 
(Same discourse holds for the $C_\mu$'s 
of \sref{sec:omtlaint_proc} and \sref{sec:multiomt_proc}.)
In the subsequent calls such clauses are initially inactive,
but they can be activated as soon as the current minimum, namely $\minvalue'$, 
becomes smaller or equal than \minvalue and the novel clause
$(\cost<\minvalue')$ is learned, so that $(\cost<\minvalue)$ can be
\T-propagated or $(\neg(\cost<\minvalue')\vee(\cost<\minvalue))$ can be
\T-learned. This allows for reusing lots of previous search.     
% \end{remark}

\section{Experimental Evaluation}
\label{sec:eval}

We have extended \optimathsat \cite{st-ijcar12,st_tocl14} by
implementing the advanced and truncated B\&B
 \omlaplus{} procedures described in
\sref{sec:omtlaint_proc}. 
On top of that, we have implemented our
techniques for multi-objective OMT
(\sref{sec:multiomt_proc}) 
{---including the lexicographic combination---} 
and incremental OMT
(\sref{sec:incomt_proc}).
Then, we have investigated empirically the efficiency of our 
 new procedures by conducing
two different experimental evaluations,  respectively on \omla{} 
(\sref{sec:omtlaint_eval}) and 
 on  multi-objective and incremental
\omlarat (\sref{sec:multiomt_eval}).
All tests in this section were executed on two identical
\textit{8-core 2.20Ghz Xeon} machines with $64$ GB of RAM and running
Linux with 3.8-0-29 kernel, with an enforced timeout of 1200 seconds.

For every problem in this evaluation, the correctness of the minimum
costs  found by \optimathsat and its competitor tools,
namely ``\minvalue'', have
been cross-checked with the SMT solver \zthree, by checking both
the inconsistency of $\varphi \wedge (\cost < \minvalue)$ and the
consistency of $\varphi \wedge (\cost = \minvalue)$.  In all tests,
when terminating, all tools returned the correct results.
To make the experiments reproducible, the full-size plots,
a Linux binary of \optimathsat, the input \omt problems, and the results 
are available.~\footnote{%
\url{http://disi.unitn.it/~trentin/resources/tacas15.tar.gz}; 
\bclt is available at \url{http://www.lsi.upc.edu/~oliveras/bclt.gz};
\symba is available at \url{https://bitbucket.org/arieg/symba/src}; 
\nuz is available at \url{http://rise4fun.com/z3opt}.}

\subsection{Evaluation of \omla procedures}
\label{sec:omtlaint_eval}
Here we consider three different configurations of \optimathsat based
on the search schemas (linear vs. binary vs. adaptive, denoted
respectively by \textsc{``-lin''}, \textsc{``-bin''} and
\textsc{``-ada''}) presented in \sref{sec:background_omt}; the
adaptive strategy dynamically switches the search schemas between linear
and binary search, based on the heuristic described in
\cite{st_tocl14}. We run \optimathsat both with the advanced and
truncated branch\&bound minimization procedures for \la{} presented
in \sref{sec:omtlaint_proc}, denoted respectively by \textsc{``-adv''}
and \textsc{``-trn''}.

In order to have a comparison of \optimathsat{} with {both \nuz and}
\bclt, in this experimental evaluation we restricted our focus on 
\omlaint{} only. Here we do not consider \symba, since it does not support 
\omlaint{}.
We used as benchmarks a set of $544$ problems derived from \smt-based
Bounded Model Checking and K-Induction on parametric problems, generated 
via the \sal{} model checker.~\footnote{\url{http://sal.csl.sri.com/}.}

The results of this evaluation are shown in Figure
\ref{fig:omlaint_stats}.  By looking at the table, we
observe that the best \optimathsat{} configuration  on these
benchmarks is \textsc{-trn-ada}, which uses the truncated
branch\&bound approach within the \laintminimize procedure with
adaptive search scheme. We notice that the differences in
performances among the various configurations of \optimathsat are
small on these specific benchmarks.

Comparing the \optimathsat versions against \bclt and \nuz, we notice that 
\optimathsat {and \nuz} solve all input
formulas regardless of their configuration, 
{\nuz having better time performances,} 
whilst \bclt
timeouts on $44$ problems. 

% \ignoreinshort{%
% \RSTODO{AMPLIARE\\} 
% From the scatter-plot on the right,
% where we compare our best-performing \optimathsat{} configuration
% against \bclt, we notice that as soon as the problems become more difficult
% \optimathsat performs increasingly better than \bclt.
% }

\begin{figure}[t]

\begin{minipage}[t]{1\linewidth}
\vspace{0pt}
\centering
\begin{tabular}{|l||r|r|r|r|r|r|}
\hline
\textbf{Tool:} & \textbf{\#inst.}	& \textbf{\#solved}	
	& \textbf{\#timeout} & \textbf{time}\\
\hline
\bclt & 544 & 500 & 44 & 93040\\
\hline
\nuz{} & 544 & 544 & 0 & 36089\\
\hline
OptiM.-adv-lin & 544 & 544 & 0 & 91032\\
OptiM.-adv-bin & 544 & 544 & 0 & 99214\\
OptiM.-adv-ada & 544 & 544 & 0 & 88750\\
\hline
OptiM.-trn-lin & 544 & 544 & 0 & 91735\\
OptiM.-trn-bin & 544 & 544 & 0 & 99556\\
OptiM.-trn-ada & 544 & 544 & 0 & 88730\\
\hline
\end{tabular}
\end{minipage}

% \ignoreinshort{
% \begin{minipage}[t]{0\linewidth}
% \vspace{-0pt}
% \centering
% \scalebox{0.33}{\includegraphics{bclt/bclt_optimathsat-trn-ada--unbounded.pdf}}
% \end{minipage}

% \begin{minipage}[t]{1.45\linewidth}
% \vspace{-145pt}
% \centering
% \scalebox{0.33}{\includegraphics{bclt/muz3_optimathsat-trn-ada--unbounded.pdf}}
% \end{minipage}
% }

\ignoreinshort{
\begin{tabular}{ll}
\centering
 \scalebox{0.25}{\includegraphics{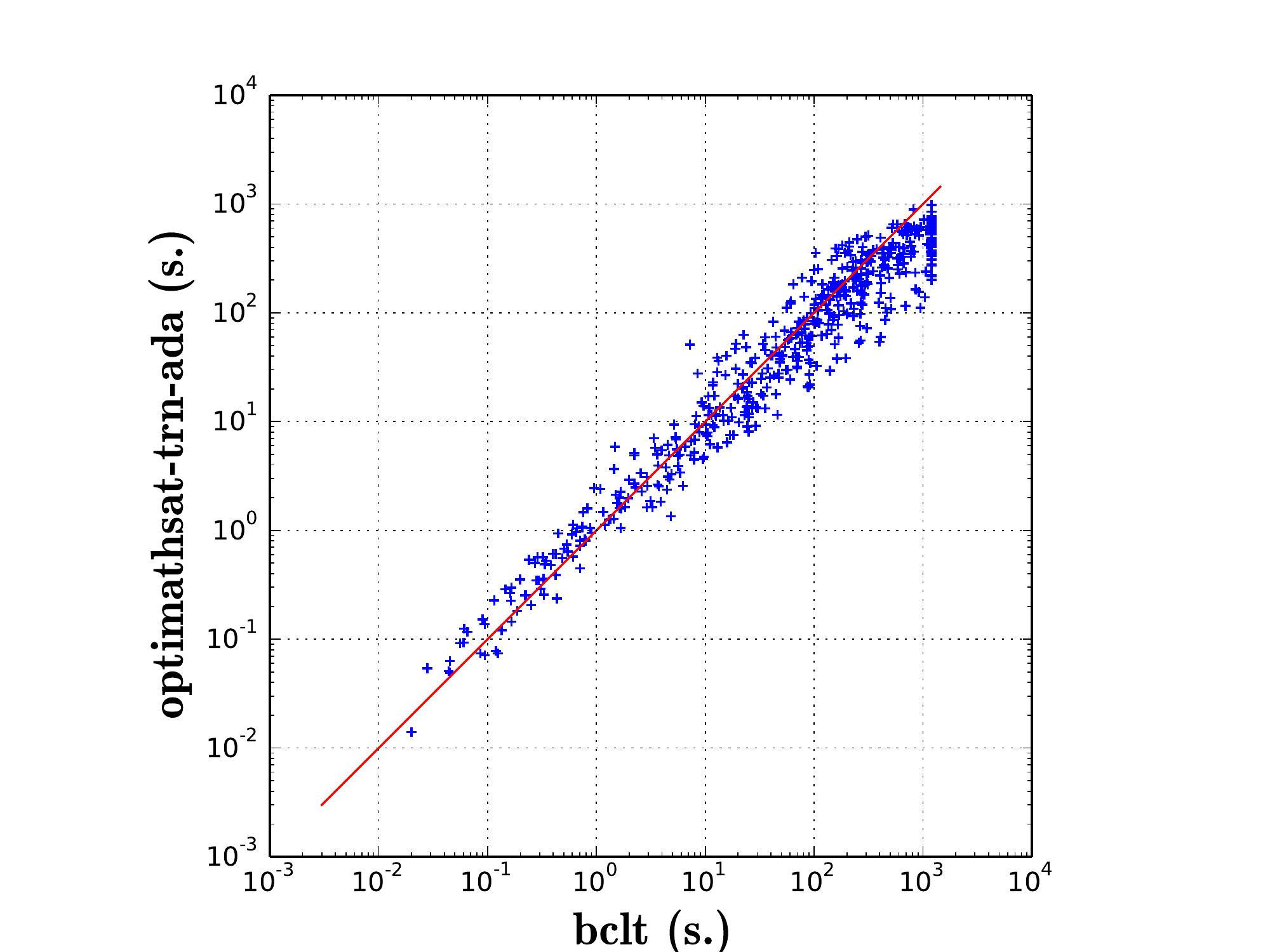}}
&
\centering
 \scalebox{0.25}{\includegraphics{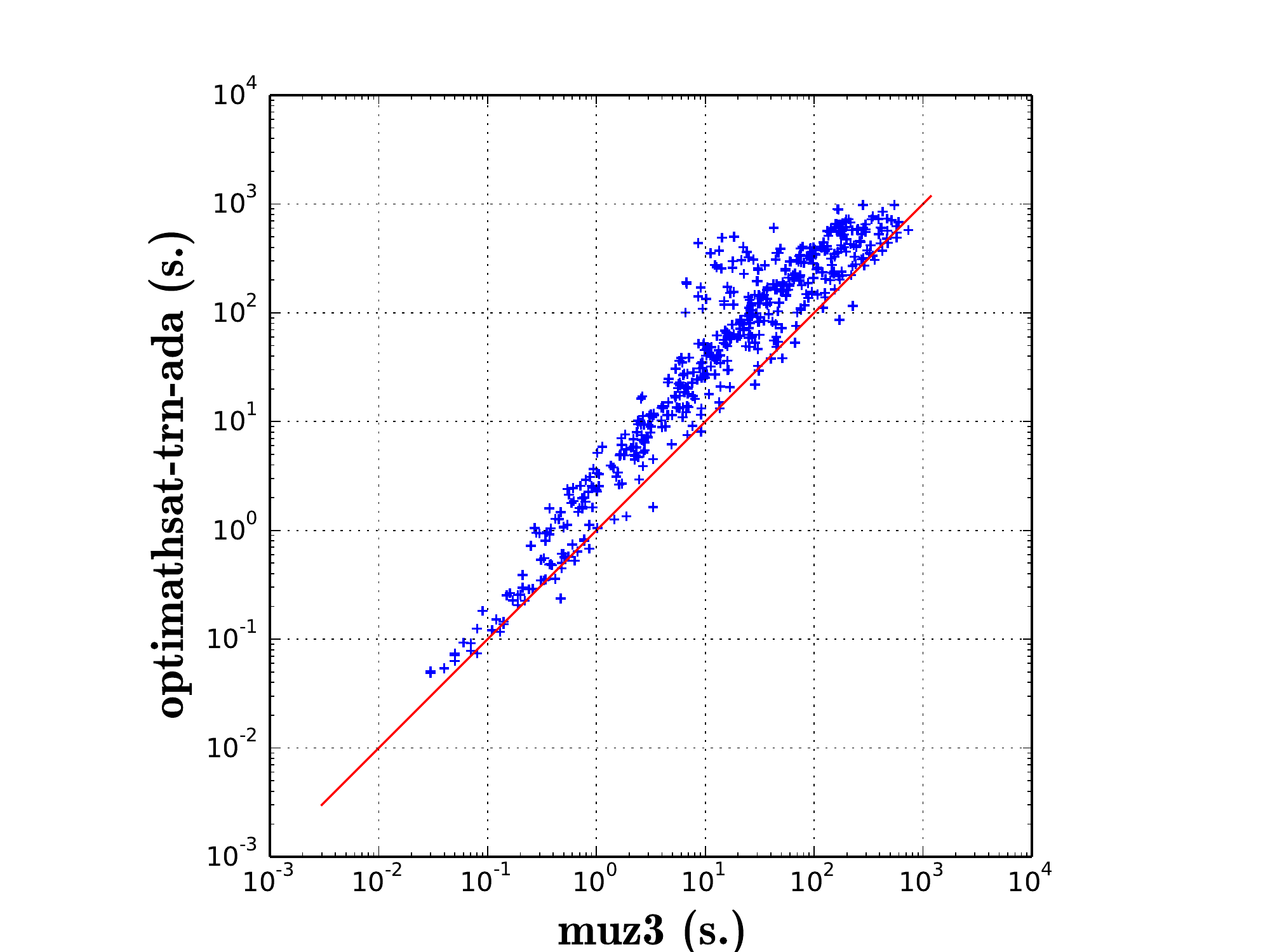}}
%&
% \centering
%  \scalebox{0.25}{\includegraphics{bclt/muz3_optimathsat-trn-ada_unsat_norm--unsat.pdf}}
\end{tabular}
}

\caption{\label{fig:omlaint_stats} 
{A table comparing the performances
of \bclt, \nuz and different configurations of \optimathsat   on
Bounded Model Checking problems.
\ignoreinshort{
Scatterplots: pairwise comparisons between 
\optimathsat-trn-ada and \bclt (left) and \nuz (right).
}}
}
\end{figure}

\subsection{Evaluation of Incremental and Multiple-objective OMT}
\label{sec:multiomt_eval}

% Also, so far neither \symba nor
% \bclt handle optimization over multiple theories.
%  Moreover, the only other \omt tool featuring multi-objective optimization is
% \symba, which handles only \omlarat{} and it currently does not support strict
% inequalities inside the input formulas. 
% %
% Thus, in order to test the efficiency of our multiple-objective OMT
% approach, we compared various versions of \optimathsat against the the two best-performing
% versions of \symba presented in \cite{li_popl14}, namely \symba{}(100)
% and \symba{}(40)+\optzt.

{%
As mentioned in Section \sref{sec:intro}, so far 
\bclt does not 
feature  multi-objective OMT, and neither \symba nor
\bclt implement incremental \omt. 
\ignore{%
This forces us to evaluate the efficiency of our incremental OMT approach
by comparing an incremental version of \optimathsat against the corresponding
non-incremental one. 
Also, the only other \omt tool featuring multi-objective optimization is
\symba. 
}}
Thus, in order to test the efficiency of our multiple-objective OMT
approach, we compared three versions of \optimathsat against 
{the corresponding versions of \nuz and} 
the two
best-performing versions of \symba presented in \cite{li_popl14},
namely \symba{}(100) and \symba{}(40)+\optzt.

So far \symba handles only \omlarat{}, without combinations with other
theories.  Moreover, it currently does not support strict inequalities
inside the input formulas.
Therefore for both comparisons we used as benchmarks
the multiple-objective problems which were proposed in \cite{li_popl14} to evaluate
\symba, which were generated from a set of C programs used in the 2013
SW Verification
Competition.~\footnote{\url{https://bitbucket.org/liyi0630/symba-bench}.}
%
%\begin{remark}
  Also, \symba computes both the minimum and the maximum value for each
  $\cost$ variable, and there is no way of restricting its
  focus only on one direction. Consequently, in our tests we have
  forced also \optimathsat {and \nuz} to both minimize and maximize each
  objective. (More specifically, they had to minimize both $\cost_i$ and
  $-\cost_i$, for each $\cost_i$.)
%\end{remark}

%  \begin{rs}
%For each problem %$\tuple{\vi,\calc}$ 
We tested three different configurations of \nuz and \optimathsat:

\begin{itemize}
\item {\sc singleobjective}: each tool
  is run singularly on the single-objective problems
$\tuple{\vi,\cost_i}$ and \tuple{\vi,-\cost_i} for each $\cost_i$,
and the cumulative time is taken;

\item {\sc incremental}: as above, using the incremental
  version of each tool, each time popping
  the definition of the 
  previous \cost and pushing the new one;

\item {\sc multiobjective}: each tool is run in
  multi-objective mode with  $\bigcup_i\{\cost_i,-\cost_i\}$.
\end{itemize}    
%  \end{rs}

% \ignore{
% For each problem $\tuple{\vi,\calc}$ 
% we tested three different configurations of \optimathsat:

% \begin{itemize}
% \item \optimathsat{\sc-singleobjective}: standard \optimathsat
%   is run singularly on the single-objective problems
% $\tuple{\vi,\cost_i}$ and \tuple{\vi,-\cost_i} for each $\cost_i$,
% and the cumulative time is taken;

% \item \optimathsat{\sc-incremental}: as above, using the incremental
%   version of \optimathsat (\sref{sec:incomt_proc}), each time popping
%   the definition of the 
%   previous \cost and pushing the new one;

% \item \optimathsat{\sc--multiobjective}: \optimathsat is run in
%   multi-objective mode (\sref{sec:multiomt_proc}) with  $\bigcup_i\{\cost_i,-\cost_i\}$.
% \end{itemize}    
% }

\begin{figure}[t] 
\center
	\scalebox{0.5}{
		\includegraphics{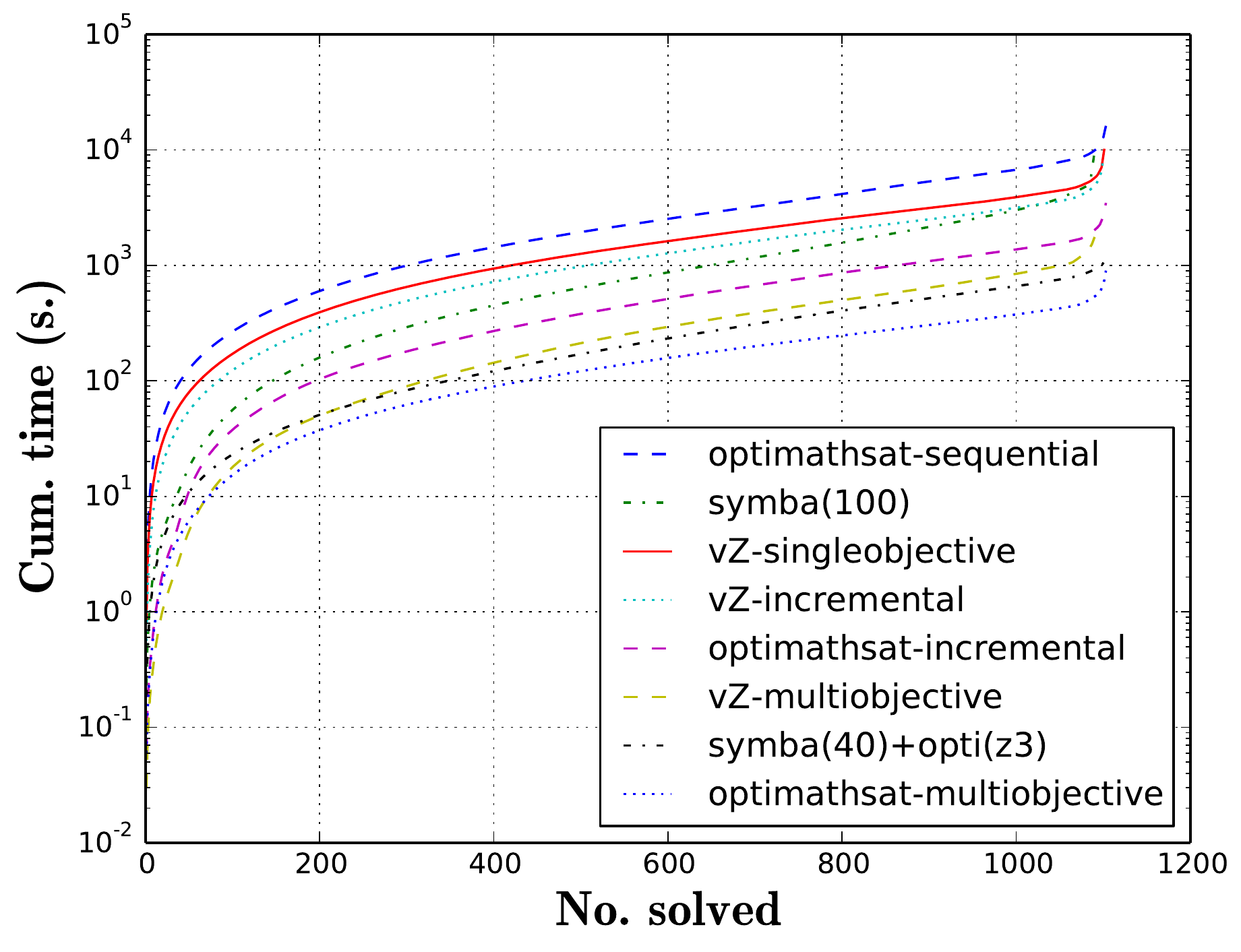}
	} 
	% \caption{\label{fig:multiomt_cactus} cactus plot comparing \optimathsat vs \symba on the software verification set of benchmarks.}
% \end{figure} 

% \begin{table}[t]
\centering
\begin{tabular}{|l||c|c|c|c|r|}
\hline
\textbf{Tool:} & \textbf{\#inst.}	& \textbf{\#solved}	
	& \textbf{\#timeout} & \textbf{time}\\
\hline
\symba{}(100) & 1103 & 1091 & 12 & 10917\\
\symba{}(40)+\optzt & 1103 & 1103 & 0 & 1128\\
\hline
\nuz{}-multiobjective & 1103 & 1090 & 13 & 1761\\
\nuz{}-incremental & 1103 & 1100 & 3 & 8683\\
\nuz{}-singleobjective & 1103 & 1101 & 2 & 10002\\
\hline
% optimathsat-1.1 (old)
%optimathsat-multiobjective & 1103 & 1103 & 0 & 939\\
%optimathsat-incremental & 1103 & 1103 & 0 & 3472\\
%optimathsat-singleobjective & 1103 & 1103 & 2 & 15305\\
% optimathsat-1.2.1 (new)
optimathsat-multiobjective & 1103 & 1103 & 0 & 901 \\
optimathsat-incremental & 1103	 & 1103 & 0 & 3477 \\
optimathsat-singleobjective & 1103 & 1103 & 0 & 16161 \\
\hline
\end{tabular}
\vspace{1ex}
\caption{\label{tab:multiomt_stats}Comparison of different versions of \optimathsat and \symba  on the SW
  verification problems in \cite{li_popl14}. (Notice the logarithmic
  scale of  the vertical axis in the cumulative plots.)}
\end{figure}

\begin{figure}[t]
	\small
	\centering
	\footnotesize
	\begin{tabular}{ccccc}
		\includegraphics[scale=0.25]{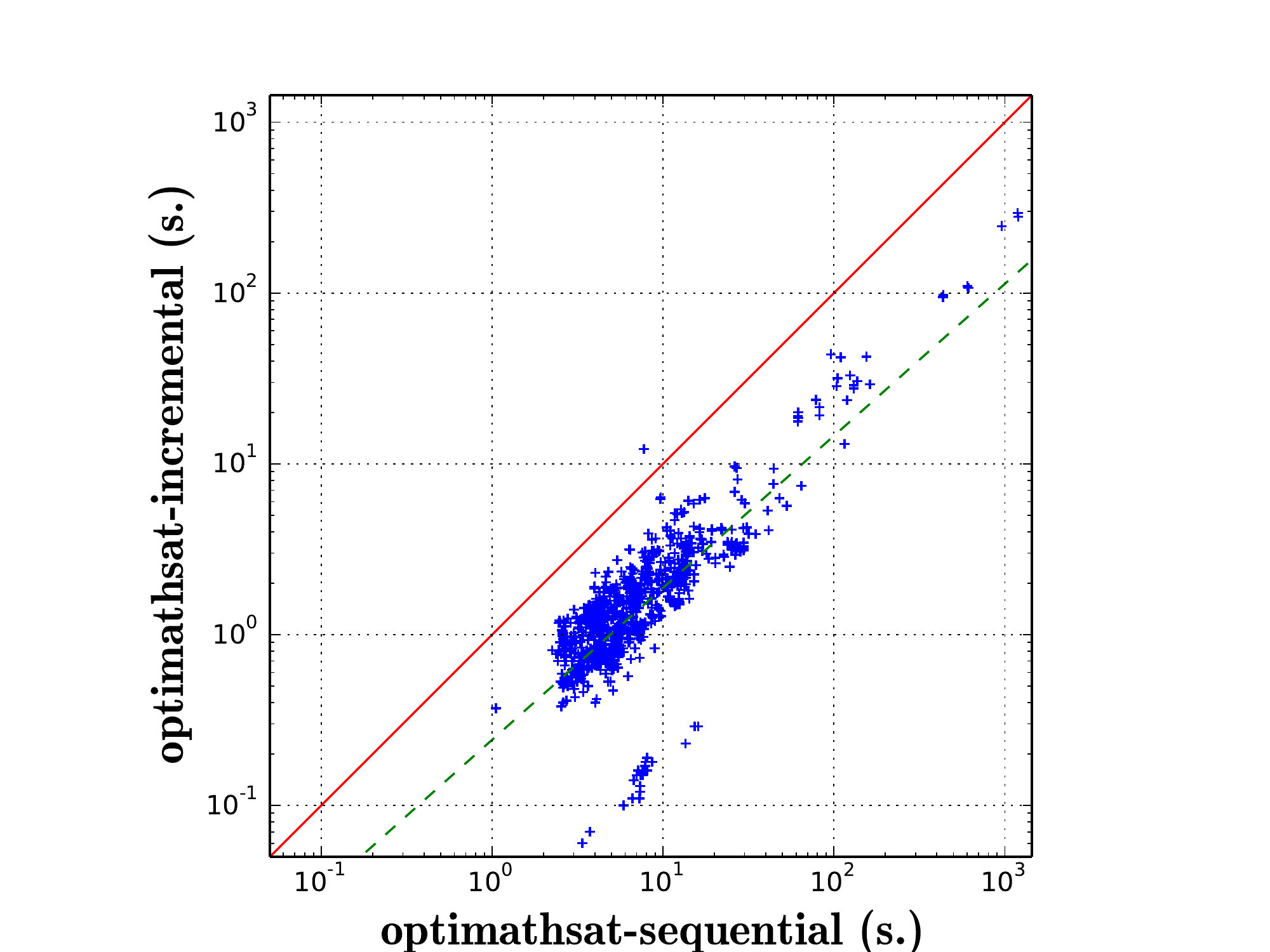}
&
\hspace{-8.5ex}
&
		\includegraphics[scale=0.25]{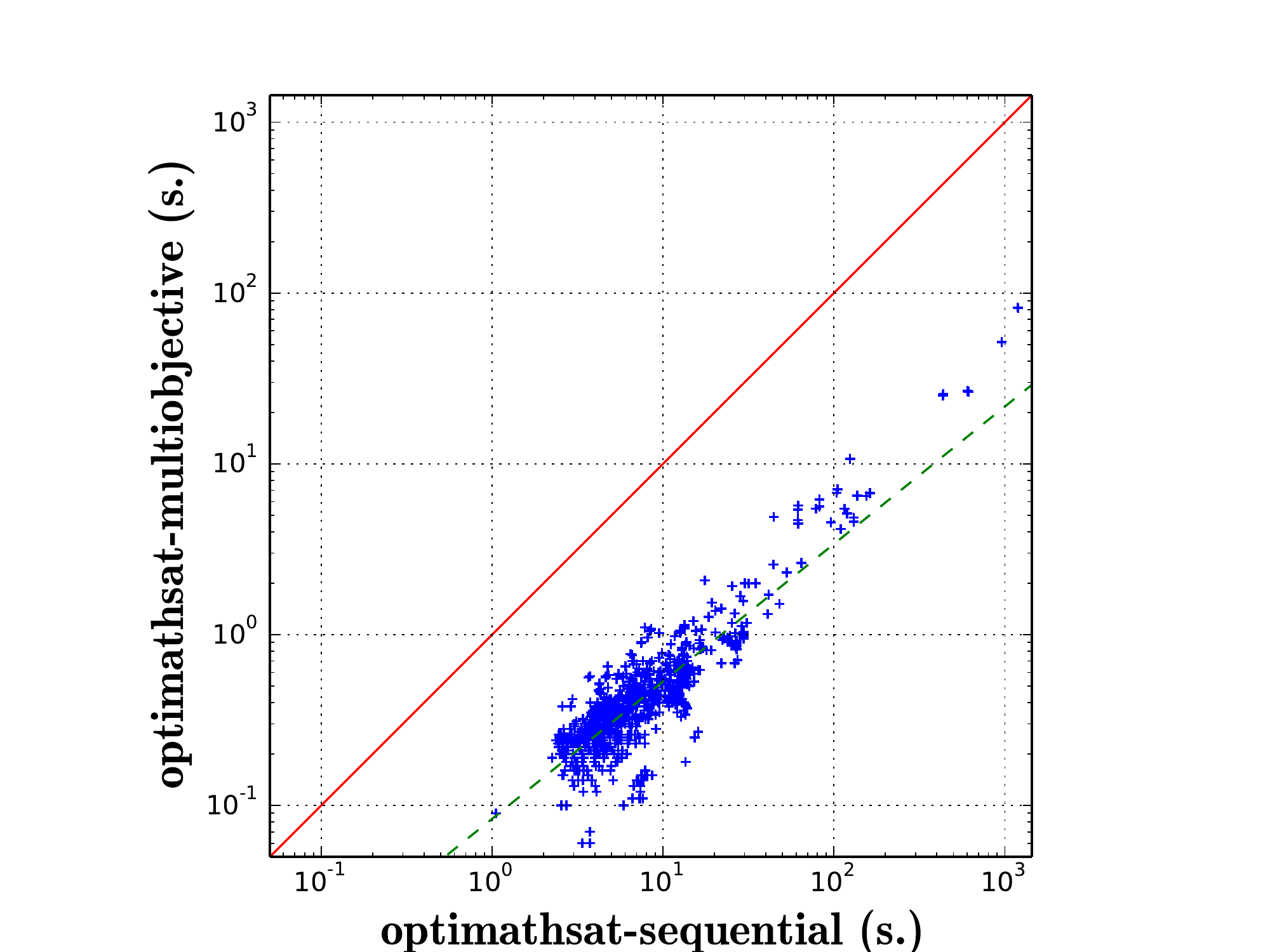}
&
\hspace{-8.5ex}
&
		\includegraphics[scale=0.25]{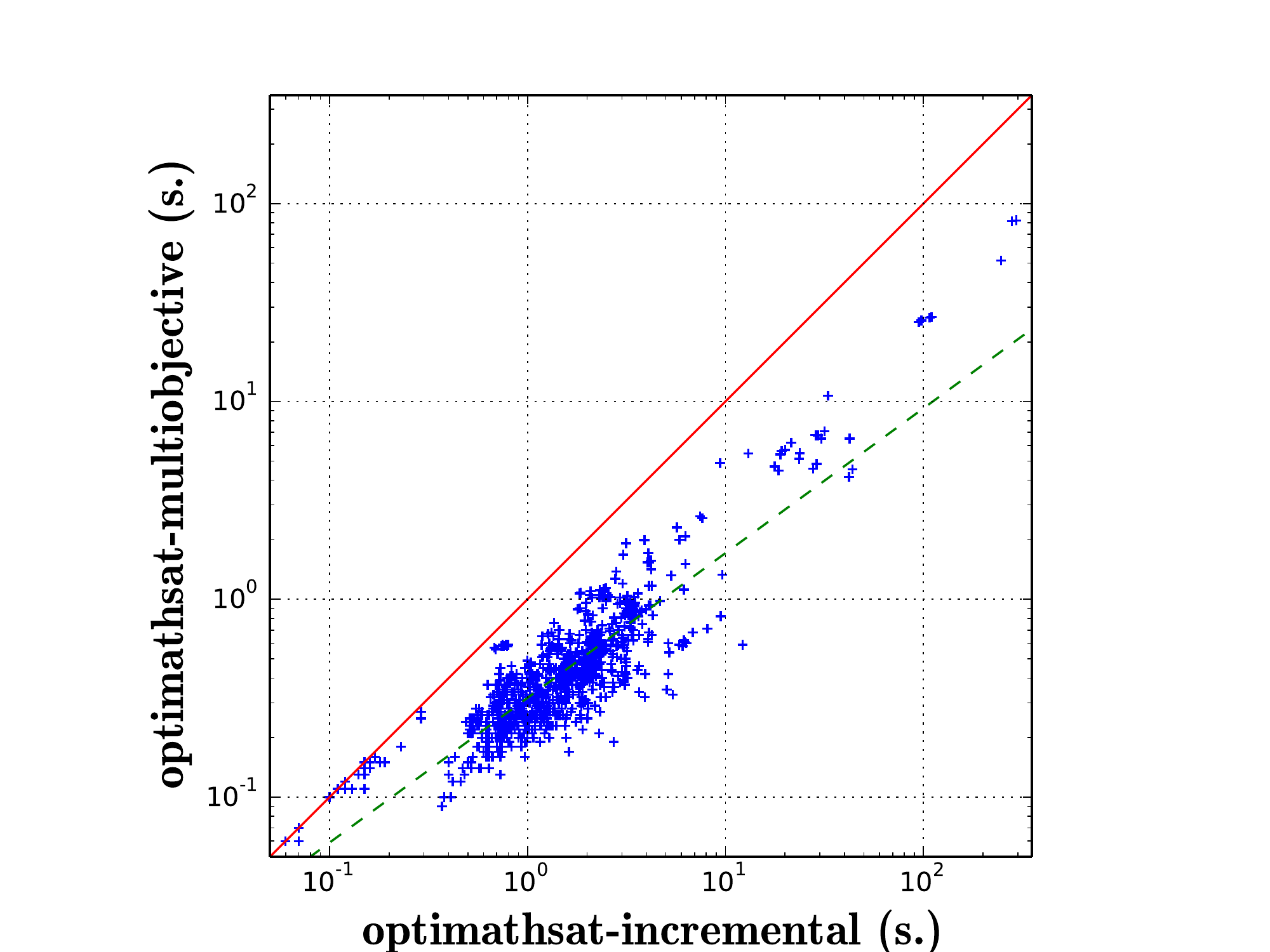}
\\
		\includegraphics[scale=0.25]{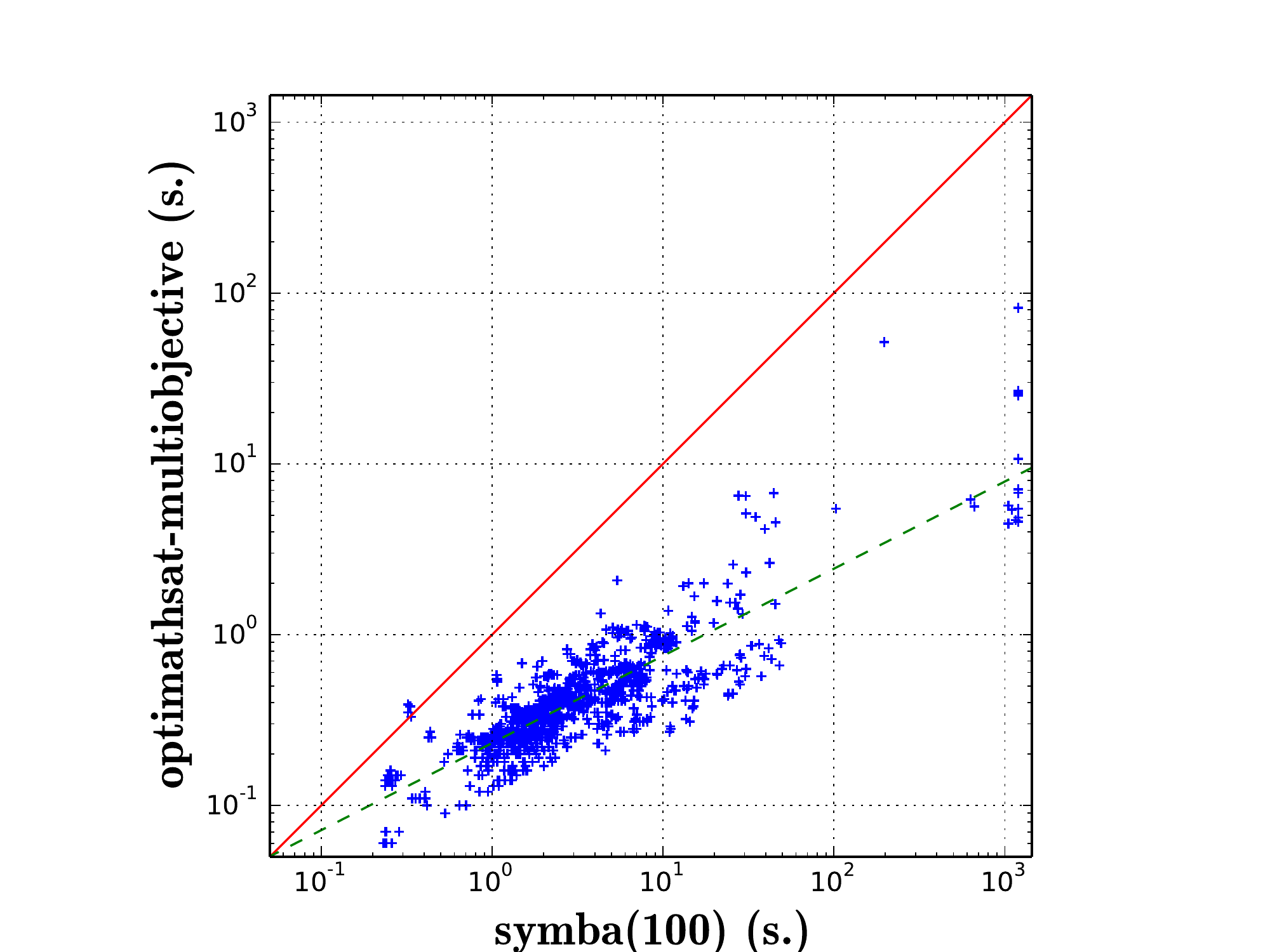}
&
\hspace{-8.5ex}
&
		\includegraphics[scale=0.25]{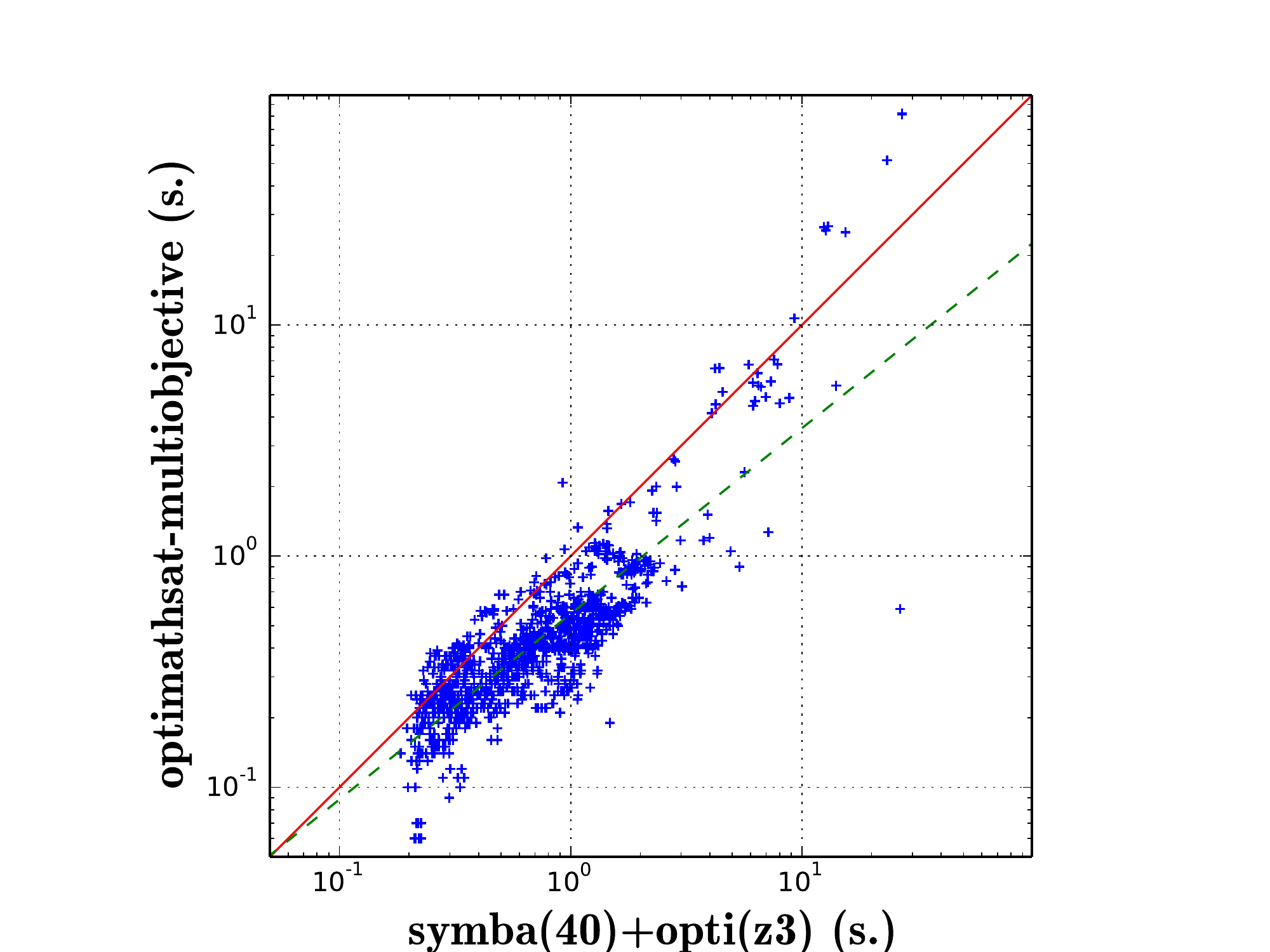}
&
\hspace{-8.5ex}
&
		\includegraphics[scale=0.25]{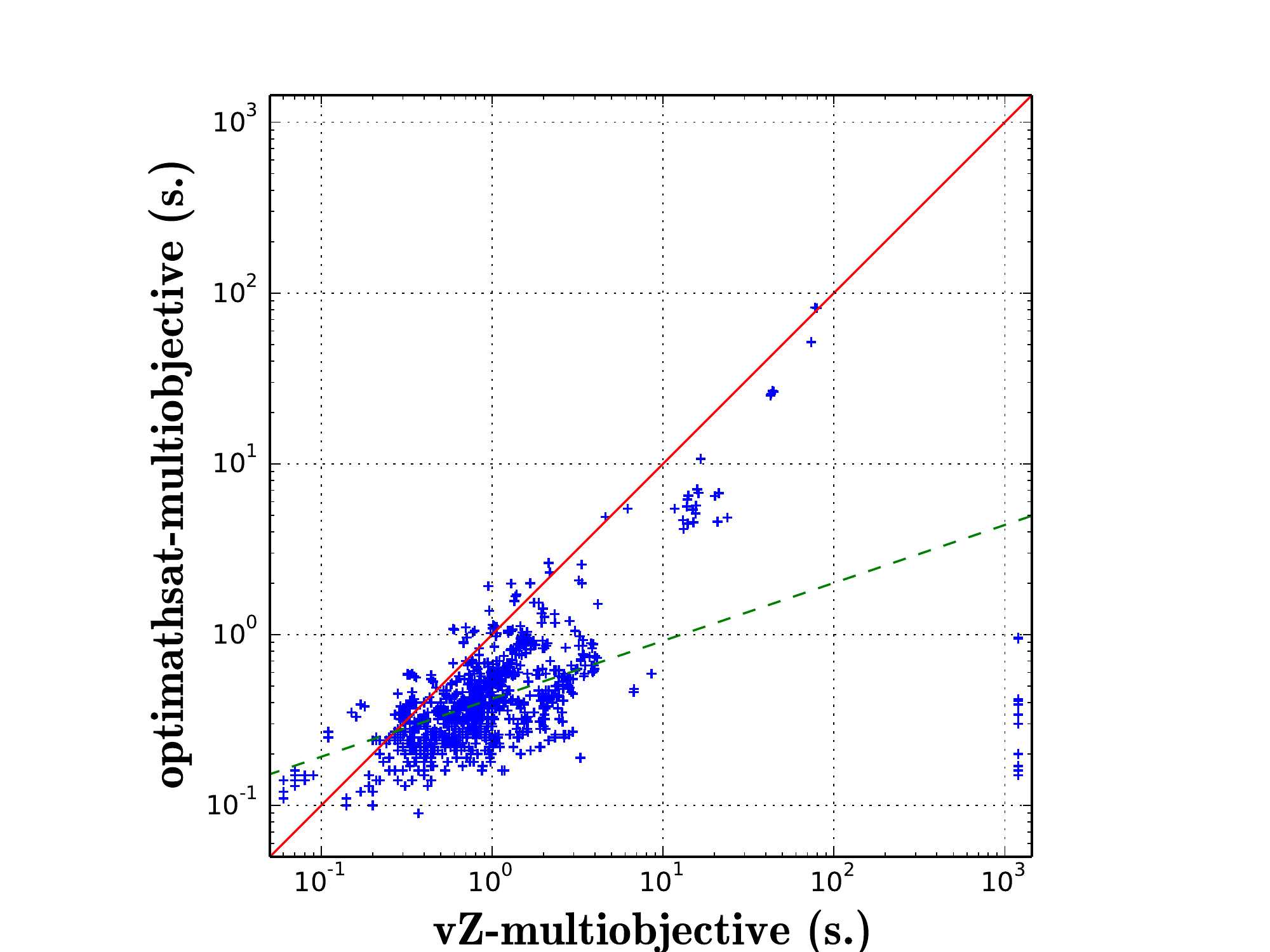}
\\
% NEW NORMALIZED
		\includegraphics[scale=0.25]{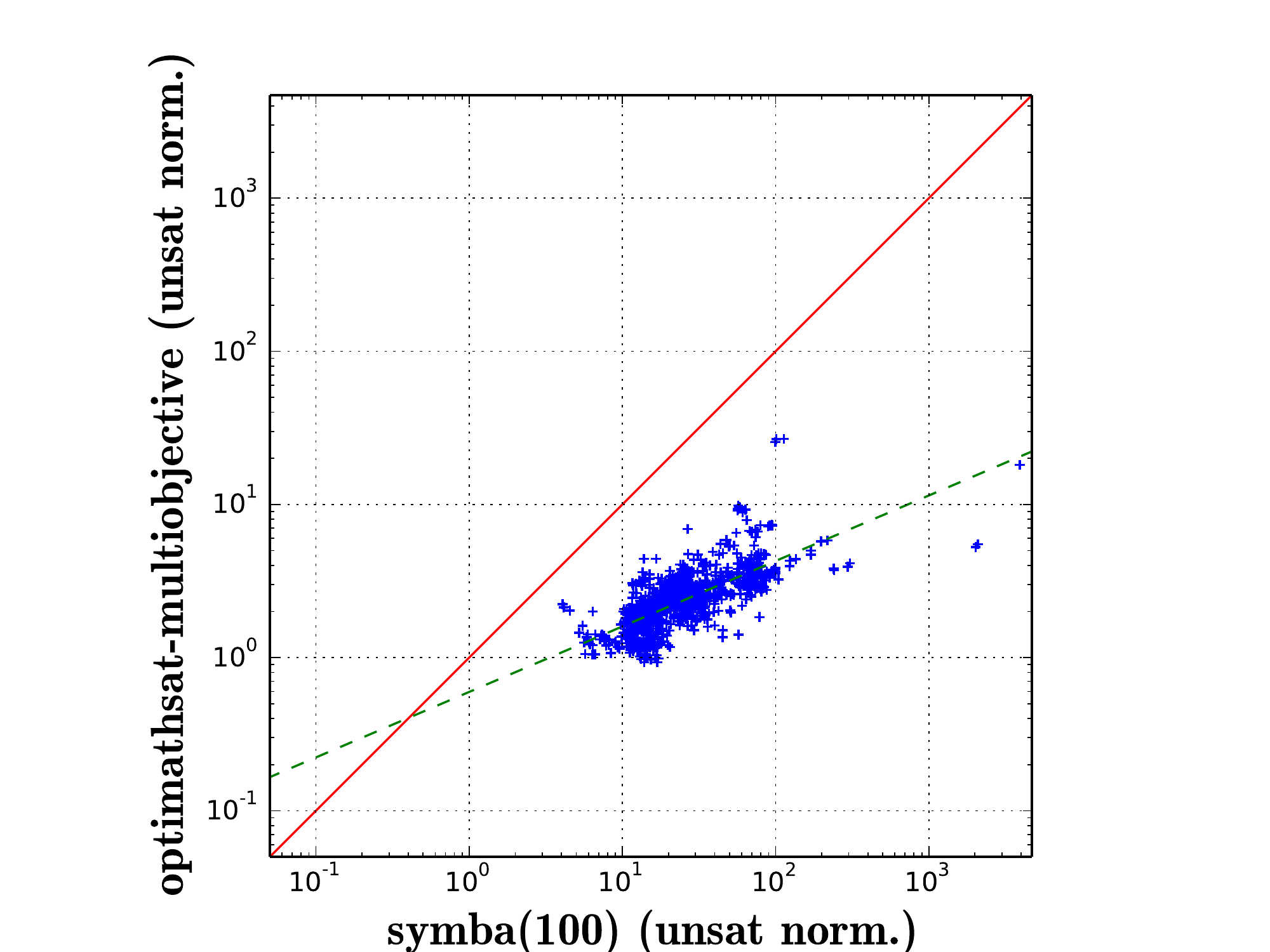}
&
\hspace{-8.5ex}
&
		\includegraphics[scale=0.25]{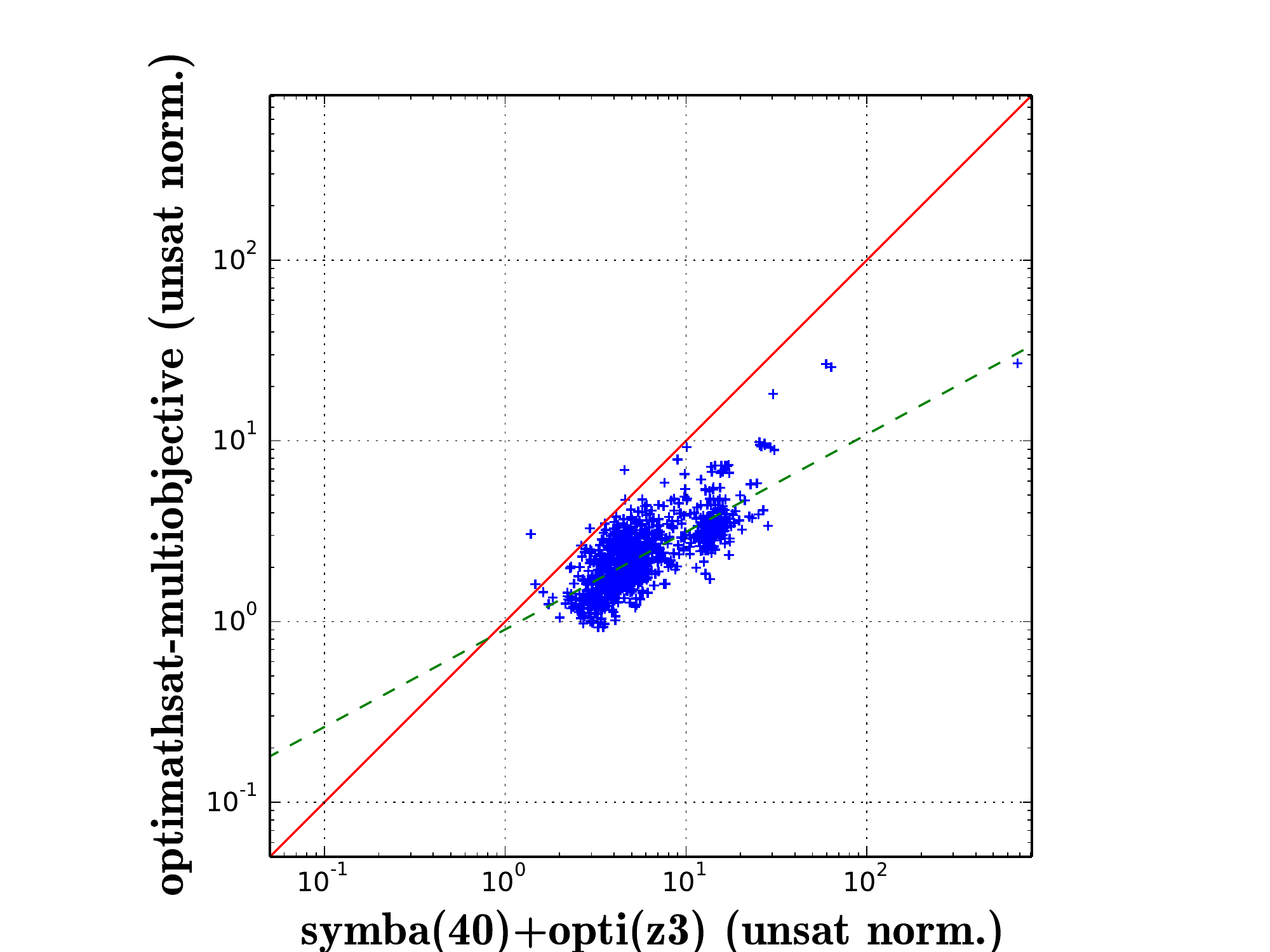}
&
\hspace{-8.5ex}
&
		\includegraphics[scale=0.25]{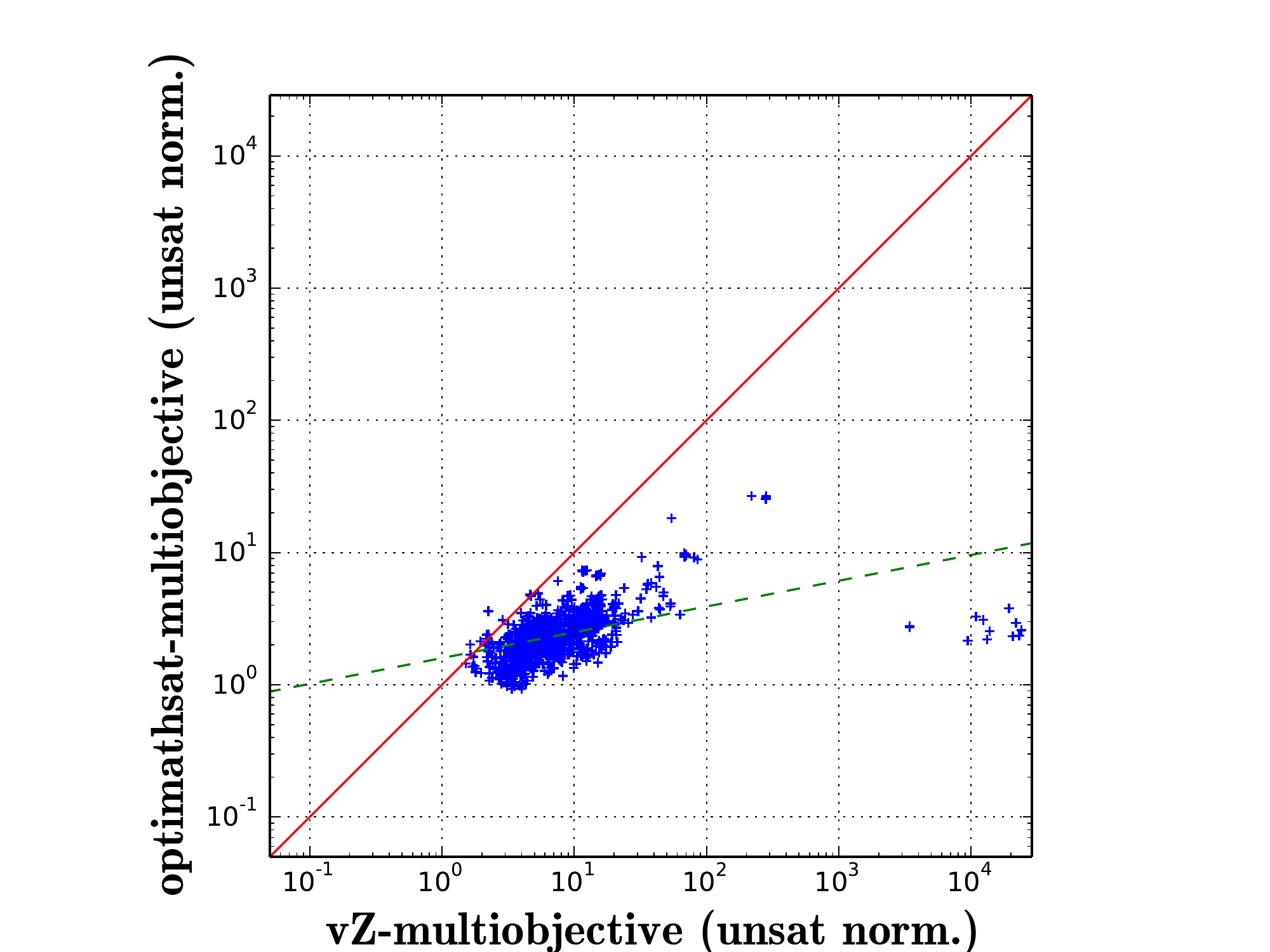}
\\
[+2ex]
	\end{tabular}
	\caption{\label{fig:multiomt_stats} 
{
First row: pairwise comparisons between different versions of
\optimathsat.
Second row: pairwise comparisons between \optimathsat{\sc-multiobjective}, the two
versions of \symba and \nuz{\sc-multiobjective}.
Third row: ``normalized'' version of the plots in the second row.
\ignore{The green dots identify unsatisfiable formulas,
whereas the blue ones identify the satisfiable ones.}
}
}
\end{figure}

Figure~\ref{tab:multiomt_stats} provides the cumulative plots and the 
global data of the performance of all procedures under test, whilst
Figure~\ref{fig:multiomt_stats} reports pairwise comparisons.

We first compare the different versions of \optimathsat 
(see Figure~\ref{tab:multiomt_stats} and the first row of
Figure~\ref{fig:multiomt_stats}). 
%
% In the first row of Figure \ref{fig:multiomt_stats} we examine in
% detail the optimization techniques presented in this paper and freshly
% implemented in \optimathsat: incrementality (\sref{sec:incomt_proc})
% and multi-objective optimization (\sref{sec:multiomt_proc}). 
%
By
looking at Figure~\ref{tab:multiomt_stats} and at the top-left plot in
Figure \ref{fig:multiomt_stats}, we observe a uniform and relevant 
speedup when passing from non-incremental to incremental OMT. 
 This is explained by the possibility of reusing learned clauses 
from one call to the other, saving thus lots of search, as explained in \sref{sec:incomt_proc}.

By
looking at Figure~\ref{tab:multiomt_stats} and at the top-center plot in
Figure \ref{fig:multiomt_stats}, we observe a uniform and drastic 
speedup in performance --about one order of magnitude-- 
when passing from single-objective to 
multiple-objective OMT. We also notice (top-right plot in
Figure \ref{fig:multiomt_stats}) that this performance is
significantly better than that obtained with incremental OMT. 
Analogous considerations hold for \nuz.

We see two main motivations for this improvement in performance with
our multiple-objective OMT technique:
first, every time a novel truth assignment is generated, the value of
many cost functions can be updated, sharing thus lots of Boolean
and \larat search; second, the process of certifying that there is no
better solution, which typically requires a significant part of the
overall OMT search \cite{st_tocl14}, here is executed only once.

% We observe that our technique for
% multi-objective optimization is even more drastically better, since it
% pays off up to an order of magnitude with respect to single-objective
% optimization. This data is confirmed by looking at table
% \ref{tab:multiomt_stats}, and notice that \optimathsat requires less
% $\frac{1}{15}$ of time when we use our technique for multi-objective
% optimization rather than separately optimize each $\cost$ with the
% single-objective technique.  In this regard, the result that we obtain
% is no surprise: to use \optimathsat in single-objective mode we had to
% split our original set of $1103$ in $62.110$ distinct optimization
% sub-problems. At last, in the top-right plot we compare the
% multi-objective technique with the incremental approach and observe
% that, when it is possible, it is uniformly better to optimize multiple
% objectives in a single run than incrementally.

%\begin{rs}
In the second row of Figure \ref{fig:multiomt_stats} we compare the
performances of \optimathsat{\sc-multi-objective} against  the
two versions of \symba and \nuz{\sc-multi-objective}. 
% From Figure~\ref{tab:multiomt_stats} and the
% bottom-left plot of Figure~\ref{fig:multiomt_stats} 
We observe that
multi-objective \optimathsat  performs much better than the default
configuration of \symba, and significantly better than both
\symba{}(40)+\optzt and \nuz{\sc-multi-objective}.

We have also wondered how much the relative performances of
\optimathsat, \symba and \nuz depend on the relative efficiency of their
 underlying SMT solvers: \mathsatfive for
\optimathsat{} and  
\zthree for \symba and \nuz. 
Thus we have run both \mathsatfive and \zthree on the
set of problems $\varphi \wedge (\cost < \minvalue)$ derived from the
original benchmarks, and used their timings  to divide the respective
\optimathsat{} and \symba{}/\nuz execution time values.%
\footnote{That is, each value represents the time taken by each \omt
  tool on \tuple{\vi,\cost_i} divided by the time taken by its
  underlying SMT solver to solve $\varphi \wedge (\cost < \minvalue)$.}
These ``normalized'' 
results, which are shown in the bottom row of
Figure~\ref{fig:multiomt_stats}, seem to suggest that the
better performances of \optimathsat are not due to better performances
of the underlying SMT solver. 
%\end{rs}

% TODO{METTERE A POSTO QUANTO SEGUE O ELIMINARLO?}
% Overall, we notice from Figure that the comparison among \optimathsat and
% \symba+\optzt shows that the former manages to be up to $16.71\%$ more
% efficient when run we adopt our multi-objective technique presented in
% \sref{sec:multiomt_proc}.  A better comparison among \optimathsat
% different techniques for optimization and the two versions of \symba
% that we have tested can be appreciated by looking at Figure
% \ref{fig:multiomt_cactus}, where we compare the cumulative solving
% time wrt the number of problems solved.

% This improvement has two causes: on one side the \smt{} optimization
% search can update multiple costs values for each truth assignments it
% generates, on the other side it can can amortize the certification
% time - which is known to generally take up to $50\%$ of search time -
% across a larger number of $\cost$ variables.

% \section{Conclusions and Future Work}
% \label{sec:concl} 
% \input{concl}

\FloatBarrier
 
%\ignoreinshort{\newpage}
%\pagenumbering{roman}
\bibliographystyle{abbrv} 
\bibliography{pt_refs,rs_refs,rs_ownrefs,rs_specific,sathanbook}

\begin{thebibliography}{10}

\bibitem{AudemardBCS05}
G.~Audemard, M.~Bozzano, A.~Cimatti, and R.~Sebastiani.
\newblock {Verifying Industrial Hybrid Systems with MathSAT}.
\newblock In {\em Proc. BMC 2004}, volume 119 of {\em ENTCS}. Elsevier, 2005.

\bibitem{acks_forte02}
G.~Audemard, A.~Cimatti, A.~Korni{\l}owicz, and R.~Sebastiani.
\newblock {SAT-Based Bounded Model Checking for Timed Systems}.
\newblock In {\em Proc. FORTE'02.}, volume 2529 of {\em LNCS}. Springer, 2002.

\bibitem{BSST09HBSAT}
C.~Barrett, R.~Sebastiani, S.~A. Seshia, and C.~Tinelli.
\newblock {\em Satisfiability Modulo Theories}, chapter~26, pages 825--885.
\newblock In Biere et~al. \cite{HandbookOfSAT2009}, February 2009.

\bibitem{HandbookOfSAT2009}
A.~Biere, M.~J.~H. Heule, H.~van Maaren, and T.~Walsh, editors.
\newblock {\em Handbook of Satisfiability}.
\newblock IOS Press, February 2009.

\bibitem{bjorner_scss14}
N.~Bjorner and A.-D. Phan.
\newblock {$\nu{}Z$ - Maximal Satisfaction with Z3}.
\newblock In {\em Proc SCSS. Invited presentation.}, Gammart, Tunisia, December
  2014. EasyChair Proceedings in Computing (EPiC).
\newblock \url{http://www.easychair.org/publications/?page=862275542}.

\bibitem{bozzanobcjrrs06}
M.~Bozzano, R.~Bruttomesso, A.~Cimatti, T.~A. Junttila, S.~Ranise, P.~van
  Rossum, and R.~Sebastiani.
\newblock {Efficient Theory Combination via Boolean Search}.
\newblock {\em Information and Computation}, 204(10):1493--1525, 2006.

\bibitem{byrd_unboundedMILP}
R.~H. Byrd, A.~J. Goldman, and M.~Heller.
\newblock {Technical Note-- Recognizing Unbounded Integer Programs}.
\newblock {\em Operations Research}, 35(1), 1987.

\bibitem{cimattifgss10}
A.~Cimatti, A.~Franz{\'e}n, A.~Griggio, R.~Sebastiani, and C.~Stenico.
\newblock Satisfiability modulo the theory of costs: Foundations and
  applications.
\newblock In {\em TACAS}, volume 6015 of {\em LNCS}, pages 99--113. Springer,
  2010.

\bibitem{cgss_sat13_maxsmt}
A.~Cimatti, A.~Griggio, B.~J. Schaafsma, and R.~Sebastiani.
\newblock {A Modular Approach to MaxSAT Modulo Theories}.
\newblock In {\em SAT}, volume 7962 of {\em LNCS}, July 2013.

\bibitem{mathsat5_tacas13}
A.~Cimatti, A.~Griggio, B.~J. Schaafsma, and R.~Sebastiani.
\newblock {The MathSAT 5 SMT Solver}.
\newblock volume 7795 of {\em LNCS}. Springer, 2013.

\bibitem{dilligdma12}
I.~Dillig, T.~Dillig, K.~L. McMillan, and A.~Aiken.
\newblock {Minimum Satisfying Assignments for SMT}.
\newblock In {\em CAV}, pages 394--409, 2012.

\bibitem{demoura_cav06}
B.~Dutertre and L.~de~Moura.
\newblock {A Fast Linear-Arithmetic Solver for DPLL(T).}
\newblock In {\em CAV}, volume 4144 of {\em LNCS}, 2006.

\bibitem{eensorensson:sat2003}
N.~E{\'e}n and N.~S{\"o}rensson.
\newblock An extensible {SAT}-solver.
\newblock In {\em Theory and Applications of Satisfiability Testing (SAT
  2003)}, volume 2919 of {\em LNCS}, pages 502--518. Springer, 2004.

\bibitem{griggio-jsat11}
A.~Griggio.
\newblock {A Practical Approach to Satisfiability Modulo Linear Integer
  Arithmetic}.
\newblock {\em Journal on Satisfiability, Boolean Modeling and Computation -
  JSAT}, 8:1--27, 2012.

\bibitem{LarrazORR14}
D.~Larraz, A.~Oliveras, E.~Rodr{\'{\i}}guez{-}Carbonell, and A.~Rubio.
\newblock {Minimal-Model-Guided Approaches to Solving Polynomial Constraints
  and Extensions}.
\newblock In {\em {SAT}}, 2014.

\bibitem{li_popl14}
Y.~Li, A.~Albarghouthi, Z.~Kincad, A.~Gurfinkel, and M.~Chechik.
\newblock {Symbolic Optimization with SMT Solvers}.
\newblock In {\em POPL}. ACM Press., 2014.

\bibitem{manoliosp13}
P.~Manolios and V.~Papavasileiou.
\newblock Ilp modulo theories.
\newblock In {\em CAV}, pages 662--677, 2013.

\bibitem{MSLM09HBSAT}
J.~P. Marques-Silva, I.~Lynce, and S.~Malik.
\newblock {\em Conflict-Driven Clause Learning SAT Solvers}, chapter~4, pages
  131--153.
\newblock In Biere et~al. \cite{HandbookOfSAT2009}, February 2009.

\bibitem{nieuwenhuis_sat06}
R.~Nieuwenhuis and A.~Oliveras.
\newblock {On SAT Modulo Theories and Optimization Problems}.
\newblock In {\em SAT}, volume 4121 of {\em LNCS}. Springer, 2006.

\bibitem{nieot-jacm-06}
R.~Nieuwenhuis, A.~Oliveras, and C.~Tinelli.
\newblock {Solving SAT and SAT Modulo Theories: from an Abstract
  Davis-Putnam-Logemann-Loveland Procedure to DPLL(T)}.
\newblock {\em Journal of the ACM}, 53(6):937--977, November 2006.

\bibitem{sebastiani07}
R.~Sebastiani.
\newblock {Lazy Satisfiability Modulo Theories}.
\newblock {\em Journal on Satisfiability, Boolean Modeling and Computation,
  JSAT}, 3(3-4):141--224, 2007.

\bibitem{st_tocl14}
R.~Sebastiani and S.~Tomasi.
\newblock {Optimization Modulo Theories with Linear Rational Costs}.
\newblock To appear on ACM Transactions on Computational Logics, TOCL.
\newblock Available at
  \url{http://optimathsat.disi.unitn.it/pages/publications.html}.

\bibitem{st-ijcar12}
R.~Sebastiani and S.~Tomasi.
\newblock {Optimization in SMT with LA(Q) Cost Functions}.
\newblock In {\em IJCAR}, volume 7364 of {\em LNAI}, pages 484--498. Springer,
  July 2012.

\bibitem{st_tacas15}
R.~Sebastiani and P.~Trentin.
\newblock Pushing the envelope of optimization modulo theories with
  linear-arithmetic cost functions.
\newblock In {\em Proc. TACAS}, LNCS. Springer, 2015.

\end{thebibliography}

\end{document}